\documentclass[review]{sigchi}

\usepackage{balance}       % to better equalize the last page
\usepackage{graphics}      % for EPS, load graphicx instead 
\usepackage[T1]{fontenc}   % for umlauts and other diaeresis
\usepackage{txfonts}
\usepackage{mathptmx}
\usepackage[pdflang={en-US},pdftex]{hyperref}
\usepackage{color}
\usepackage{booktabs}
\usepackage{textcomp}

\graphicspath{{figures/}}

\usepackage{microtype}        % Improved Tracking and Kerning
\usepackage{ccicons}          % Cite your images correctly!
\usepackage{todonotes}

\usepackage{graphicx}
\usepackage{dsfont}
\usepackage{booktabs} % for professional tables

\usepackage{amsmath, amssymb, bbm}
\usepackage{nameref}
\usepackage{subcaption}
\usepackage{tabularx}

\usepackage{algpseudocode}

\usepackage{color-edits}
\addauthor[Maria]{md}{blue}
\addauthor[Alex]{ac}{purple}
\addauthor[Riccardo]{rf}{green}

\usepackage{tikz} % graphs
\usetikzlibrary{positioning,shapes.geometric} % graphs
\usetikzlibrary{decorations.pathreplacing,angles,quotes}
\usepackage{enumitem}

\usepackage{subcaption}

\renewcommand{\P}{\mathbb{P}}

\usepackage{mathtools}

\usepackage{hyperref}

\newcommand\blfootnote[1]{%
  \begingroup
  \renewcommand\thefootnote{}\footnote{#1}%
  \addtocounter{footnote}{-1}%
  \endgroup
}

\def\plaintitle{A Case for Humans-in-the-Loop: Decisions in the Presence of Erroneous Algorithmic Scores}

\def\emptyauthor{}
\def\plainkeywords{Authors' choice; of terms; separated; by
  semicolons; include commas, within terms only; this section is required.}

\makeatletter
\def\url@leostyle{%
  \@ifundefined{selectfont}{
    \def\UrlFont{\sf}
  }{
    \def\UrlFont{\small\bf\ttfamily}
  }}
\makeatother
\def\pprw{8.5in}
\def\pprh{11in}

\definecolor{linkColor}{RGB}{6,125,233}
\hypersetup{%
  pdftitle={\plaintitle},
  pdfauthor={\emptyauthor},
  pdfkeywords={\plainkeywords},
  pdfdisplaydoctitle=true, % For Accessibility
  bookmarksnumbered,
  pdfstartview={FitH},
  colorlinks,
  citecolor=black,
  filecolor=black,
  linkcolor=black,
  urlcolor=linkColor,
  breaklinks=true,
  hypertexnames=false
}

\toappear{Accepted at ACM CHI Conference on Human Factors in Computing Systems, 2020.}
\makeatletter
\def\@copyrightspace{\relax}
\makeatother

\begin{document}

\title{\plaintitle}

\numberofauthors{3}
\author{%
  \alignauthor{Maria De-Arteaga$^*$\\
    \affaddr{Heinz College\\ Machine Learning Department\\ Carnegie Mellon University}\\
    \affaddr{Pittsburgh, PA, USA}\\
    \email{mdeartea@andrew.cmu.edu}}\\
  \alignauthor{Riccardo Fogliato$^*$\\
    \affaddr{Department of Statistics and Data Science\\ Carnegie Mellon University}\\
    \affaddr{Pittsburgh, PA, USA}\\
    \email{rfogliat@andrew.cmu.edu}}\\
  \alignauthor{Alexandra Chouldechova\\
    \affaddr{Heinz College\\ Carnegie Mellon University}\\
    \affaddr{Pittsburgh, PA, USA}\\
    \email{achould@cmu.edu}}\\
}

\maketitle

\begin{abstract}
The increased use of algorithmic predictions in sensitive domains has been accompanied by both enthusiasm and concern. To understand the opportunities and risks of these technologies, it is key to study how experts alter their decisions when using such tools. In this paper, we study the adoption of an algorithmic tool used to assist child maltreatment hotline screening decisions. We focus on the question: Are humans capable of identifying cases in which the machine is wrong, and of overriding those recommendations? We first show that humans do alter their behavior when the tool is deployed. Then, we show that humans are less likely to adhere to the machine's recommendation when the score displayed is an incorrect estimate of risk, even when overriding the recommendation requires supervisory approval. These results highlight the risks of full automation and the importance of designing decision pipelines that provide humans with autonomy.

%we find evidence that humans are able to identify cases in which the machine's prediction is not correct and choose to discard the machine's suggestion, even when this requires them to obtain approval. 
\end{abstract}

% ACM Classfication


% \ccsdesc[300]{Human-centered computing~Haptic devices}

% Author Keywords

% Print the classficiation codes
% Please use the 2012 Classifiers and see this link to embed them in the text: \url{https://dl.acm.org/ccs/ccs_flat.cfm}

\section{Introduction}
\blfootnote{$^*$ Indicates equal contribution.\\

\vspace{0.05in}
Accepted at ACM CHI Conference, 2020.}

%Widespread deployment of risk assessment tools has been fueled by promises of 

Risk assessment tools are increasingly being incorporated into expert decision-making pipelines across domains such as criminal justice, education, health, and public services \cite{chouldechova2018case, kube2019allocating, kehl2017algorithms, caruana2015intelligible, smith2012predictive}.  These tools, which range from simple regression models to more complex machine learning models, distill available information on a given case into a risk score reflecting the likelihood of one or more adverse outcomes.   Bolstered by decades of research showing that statistical models outperform human experts on prediction tasks, there is widespread optimism that these tools will increase the quality of decisions~\cite{meehl1954clinical,dawes1989clinical,grove2000clinical, aegisdottir2006meta,kleinberg2017human}.   
%Optimistic takes assure this trend will increase the quality of decisions, while many others are concerned about risks of disparate treatment and compounding injustices. This conversation has led to a growing body of work that studies fairness-related properties of machine learning algorithms.  
% In view of research finding that hu
This optimism is tempered by evidence that, while humans provided with machine predictions may achieve improved performance, not only do they continue to underperform the machine predictions, but they may also uptake the information in ways that leads to increased disparities in decision outcomes across racial \cite{green2019disparate} and socioeconomic groups \cite{skeem2019impact}.  Such findings raise critical questions about the role of humans in the loop and human-machine complementarity in key societal domains.  In this work we focus on one important question in this area: Are humans capable of identifying certain cases where the machine's recommendation is wrong, and appropriately overriding the recommendation in such cases?

%\textbf{Alex: mention somewhere in the paper human subjects concerns.}

%Such work is undoubtedly important, but it tells us little about the effect of deploying risk assessment technologies if we lack the understanding of how humans react to these tools. Do they ignore the machine? Do they blindly follow it? Are they capable of identifying and correcting for its mistakes? Answers to these questions would allow us to better anticipate the effect of deploying algorithms with certain properties, and could also guide the development of novel algorithms by helping us understand what is needed. 
We analyze a real world child welfare decision making context where call workers are tasked with deciding whether a call concerning potential child neglect or maltreatment should be screened in for investigation. While in many instances the information communicated in the call may be enough for the call worker to make a determination, in other instances the information may be vague and inconclusive. In an effort to better focus resources on investigating cases where the children are at greatest risk, Allegheny County has deployed a risk assessment tool called the Allegheny Family Screening Tool (AFST) to assist call workers in their screening decisions.  The tool uses multi-system administrative data to assess the likelihood that children on the case will experience adverse child welfare events in the near future.  More information about the tool and its development can be found on the County's AFST website~\cite{afst}.

Some time after the tool was deployed it was discovered that a technical glitch had resulted in a subset of model inputs being incorrectly calculated in realtime.  This in turn resulted in misestimated risk scores being shown in some cases.  While, as we elaborate on below, the misestimation was often mild, and the shown score generally provided reasonable risk information, the glitch permits us a rare opportunity to investigate real world decision making in the presence of misestimated risk.  
%on all children and adults associated to a call  that uses historical administrative data to assess the likelihood that the  provides the workers with a risk score that predicts the probability of adverse outcomes--detailed in Section~\ref{sec:data}--from the covariates available in the data. 
%The tool was deployed in 2016, and in this paper we study whether call workers are less likely to adhere to the machine's recommendation when the risk score is miscalculated--in other words: do humans identify and override the machine's mistakes?  % their adherence to recommendations  that decisions did change, specially for high-scoring cases for which a policy dictates that a supervisor's approval is required to screen-out the call. At the same time, we find evidence that humans are less likely to adhere to the machine's recommendation when the score displayed is an incorrect estimate of risk, even in cases where overriding the recommendation requires a supervisor's approval.

%In this paper, we analyze how call-workers' decisions to screen in calls for investigation changed when t

%The results of what humans do when presented with an incorrect score rely on the analysis of a glitch in the system during deployment, which led to the scores being sometimes miscalculated during the period we study. 

Before proceeding, we pause to make an important point.  These types of technical issues are not uncommon.  
%There is often a discrepancy between features as represented in historical data used for model training and features as assessed at runtime.  In the child welfare context, for instance, one common source of discrepancy is the incomplete resolution of all persons associated to a referral.       
What is uncommon is for organizations to choose to be transparent about their occurrence. We recognize Allegheny County for their transparency and hope that this approach will become the norm in the deployment of algorithmic systems in sensitive societal domains.   %The shown scores are often equal or very close to the estimated score that should have been displayed, as discussed in Section~\ref{sec:data}. However, this makes it possible to analyze human adherence to inaccurate recommendations, especially when small differences in the score lead to a different policy of ``override approval'' being applied.

The remainder of this paper is organized as follows.  We begin with a discussion of related work in which we provide an overview of phenomena such as algorithm aversion and automation bias that come to bear on human decision making in the presence of algorithmic decision support tools.  We then describe the child welfare decision making context, the risk assessment tool deployment setup, and our available data.  Our analysis of the data begins by demonstrating that there was a marked change in workers' screening decisions in the post-deployment period.  Having established that an overall change in behavior did occur, we then investigate the extent to which call workers deviate from recommendations based on a misestimated risk score.  We show that workers are able to appropriately override the tool in many such cases.  We also probe questions of potential disparities in adherence to recommendations across racial and socio-economic groups, finding that the deployment of the tool neither significantly mitigates nor exacerbates disparities observed at the given level of analysis.  Lastly, we conclude with a discussion of human and system factors that we believe may have  contributed to the observed results, and outline opportunities for further research to better understand relevant factors.

\section{Background and Related work}
\label{sec:related}
Prior research has attempted to answer {\it whether} and {\it how} the deployment of algorithmic risk assessment tools affects users' decisions.  While many have advocated for the adoption of these tools on the basis of their superior predictive accuracy, findings are mixed on whether integrating prediction tools into decision making significantly improves decision quality. Indeed, research in the field suggests that the outcomes of decisions taken by a human aided by a decision support system are often no better than those taken by the human alone. 
% and even when they are, there are severe limitations to the effective combination of human and machine capabilities. 

Recent work has paid special attention to the introduction of risk assessment in the context of pretrial decision making in the criminal justice system. Although the integration of the risk assessment tools was, ex ante, expected to lead to a sharp and persistent decrease in incarceration rates, recent findings suggest that there is no impact at all \cite{roadblock} or find there is a decrease but of much smaller magnitude than initially hoped \cite{sloan2018effect}. There is consensus that these lackluster results are due at least in large part to the wide heterogeneity in judges' compliance with the tools' recommendations \cite{cohen2019judicial}. Notably, differential compliance has been shown to be a factor driving increased poor-rich~\cite{skeem2019impact} and black-white~\cite{stevenson2018assessing, evidencealbright} disparities in the post-deployment period. For instance,~\cite{evidencealbright} found that the increased racial gap in incarceration rates post-deployment was due both to inter-variation---judges in whiter counties showing higher compliance---and by intra-variations---overrides of low and moderate risk being more frequent for black than for white defendants.

More broadly, there are two competing tendencies that have been observed in studies of human compliance with algorithmic recommendations: \textit{algorithm aversion} and \textit{automation bias}. {\it Algorithm aversion}--the tendency to ignore tool recommendations after seeing that they can be erroneous--originates from a lack of agency \cite{lim1995judgemental,demichele2018criminal} and lack of transparency of the algorithm \cite{yeomans2017making}. Studies have shown that users will knowingly sacrifice accuracy in favor of gaining some control over the algorithm's output \cite{dietvorst2016overcoming}. Similarly, \cite{goodwin1999judgmental} reports an experiment in which humans override the machine's predictions when these are highly reliable. Users' reliance on the system is known to vary with the observed \cite{yu2016trust, yu2017user} and stated accuracy \cite{yin2019understanding} of the system.  However, even if the recommendations of more accurate systems are followed more often, agents affected by algorithm aversion may nevertheless prefer human judgment over algorithmic predictions even when evidence known to both the designer and the user clearly indicates that the algorithmic predictions are more accurate than human assessment \cite{dietvorst2015algorithm}.  %However, the reluctance to consistently follow the prediction of the tool is not fully justified, as evidence indicates that humans fail to improve on the predictions. 

Users affected by \textit{automation bias}, on the other hand, will follow tool recommendations despite available (but unnoticed or unconsidered) information that would indicate that the recommendation is wrong.  {\it Automation bias} consists of two classes of errors. Omission errors refer to instances where humans fails to detect problematic cases (or fail to act) because they were not flagged as such by the system. A prominent example is that of pilots in high-tech cockpits, who are prone to relying blindly on automated cues as a heuristic replacement for vigilant information seeking~\cite{mosier1998}. Commission errors refer to instances where humans take action on the basis of an erroneous algorithmic recommendation, failing to incorporate contradictory external information into the decision process.  In the clinical decision support context, commission errors may result in patients being subjected to unnecessary, potentially invasive testing or treatment.  
%; notably, the value of side-information has been shown to be mistakenly evaluated with humans, leading to overrides that worsen outcomes. 
%In support to these results, a recent work~\cite{green2019disparate} explores automation bias through an experiment on Amazon's Mechanical Turk. The authors find that humans consistently perform worse than the machine even when presented with the machine's predictions. In the context of decision support systems that recommend whether to take an action or intervention, commission errors can also refer to mistakes that result in unnecessary interventions, while omission errors are those where the human fails to pursue a necessary intervention.

Studies analyzing factors contributing to automation bias have found that complex tasks and time pressure may increase over-reliance on decision support~\cite{sarter2001supporting,goddard2011automation}. The users' experience level and their confidence in their own decisions have also been found to be causes of automation bias~\cite{marten2004computer,moray2000adaptive}. Social accountability has been found to reduce automation bias~\cite{skitka2000automation}, an important result when considering decision support systems used by experts with high public visibility or who are publicly elected, such as judges.  Meanwhile, studies focused on the causes of algorithm aversion have found that repeatedly seeing the algorithm make the same mistake leads to decreased reliance of the agent on the system~\cite{dietvorst2015algorithm}, while giving some control over the algorithm can counter this phenomenon~\cite{dietvorst2016overcoming}.

Automation bias and algorithm aversion are opposing phenomena.  While automation bias degrades decision quality by driving over-compliance with algorithmic recommendations, algorithm aversion does so by driving under-compliance.
%While in one case the human has blind trust on the algorithm, in the other case there is absolute distrust. 
There are two characteristics of the decision context that are indicative of which form of bias is likely to dominate: the type of task, and the level of automation. A significant portion of the literature analyzing automation bias has studied high-tech cockpits~\cite{mosier1998automation, skitka2000automation,cummings2004automation, sarter2001supporting}, while others have looked at healthcare diagnosis~\cite{marten2004computer}, and automated control systems for fault management in a thermal-hydraulic environment~\cite{moray2000adaptive}. All these are diagnostic tasks, where it can be assumed that there is a ``ground truth" that is in principle knowable to humans. Algorithm aversion has been mostly discussed in tasks of a different nature: prognostic tasks, where predictions pertain what will happen in the future. These tasks are often behavioral prediction tasks, which concern people's future actions or performance.  Examples include predicting students' performance on an MBA program using admissions data~\cite{dietvorst2015algorithm}, forecasting sales~\cite{lim1995judgemental, goodwin1999judgmental}, and predicting which defendants will be rearrested or will fail to appear for court if released~\cite{demichele2018criminal, roadblock, sloan2018effect}.  In these tasks there is an irreducible degree of uncertainty, and any predictive model, whether human or machine, is bound to make mistakes. 
%Perhaps in part due to the nature of the task, the level of automation is also different--it is higher in those cases in which automation bias has been found. 
The role of the level of automation on over-reliance in decision support systems has been further discussed in~\cite{cummings2004automation}.    % A vast majority of the literature finding automation bias has studied behaviors in high-tech cockpits~\cite{}

%levels of automation, type of task (is there a "ground thuth" or is there inherent uncertainty)

%effect of teams~\cite{skitka2000automation} crews reduce commission errors but nor omission errors.

%and they are poorly calibrated in the evaluation of the machine's and their own risk. 

Motivated by these challenges, recent work has explored approaches to characterizing human-machine complementarity in risk assessment contexts and devising approaches to combine the strengths of both. \cite{tan2018investigating} study differences in factors relied upon by human (crowdworker) recidivism predictions and those of COMPAS, a commercial risk assessment tool.  The authors identify systematic differences in human and machine predictions, but find that those differences could not be leveraged to improve predictions.   \cite{hilgard2019learning} suggest that, instead of generating predictions, the algorithm should be trained with a humans-in-the-loop to incorporate the human decision process, and should only report to the human simpler and useful representations of the features.  \cite{madras2018predict} propose a learning to defer model, where the algorithm chooses to make a prediction or defer to the human taking into account both the model's and the human's accuracy.  Lastly, focusing on a medical diagnosis task, \cite{raghu2019algorithmic} propose a method for triaging cases between full automation and focused human effort in a manner that improves diagnostic accuracy above the human and machine predictions alone. %   The negative results highlight the need of developing a new modeling framework to promote human-machine complementarity in which the humans' decision process is considered in the algorithmic pipeline. Such work is particularly relevant in light of empirical work showing that algorithms that center the decision-maker may help overcome algorithmic aversion. 
%For example, ~\cite{dietvorst2016overcoming} presents studies where humans are more likely to use and trust the algorithm if they are able to modify it, even if only minor control is given to the human. 
%A recent line of works aims to account for human decisions as part of the machine learning algorithms. \cite{kusner2019making} proposes a causal framework to account for humans' action as interference.
 % accuracy and the humandesigns a two-stage model, wherethe first stage  where in the first stage the tool can either decide or defer the decision to the human, that will take the final decision only in this case. 

%The research suggests that humans perpretrate, or even propagate, bias in the presence of an aid-decision tool.

Recent work has also studied the effect of providing explanations on adherence to algorithmic recommendations.  ~\cite{kizilcec2016much} and~\cite{nourani2019effects} show that the perceived accuracy of the system depends on the degree to which explanations are easily understandable, with very complicated explanations reducing perceived accuracy.  At the same time, explanations can also mislead users, as shown by ~\cite{lakkaraju2019fool}. In the context of Facebook's News Feed algorithm,~\cite{rader2018explanations} found that while explanations helped users understand how the system works, they did not help them in evaluating the correctness of the output. In the context of Reddit post removal,~\cite{jhaver2019does} found that providing explanations helped users better adhere to community guidelines. Explanations may also affect perceptions of justice of algorithmic decisions, as studied by~\cite{binns2018s}. Such perceptions may not only relate to the output but also to the predictors used by the system~\cite{van2019crowdsourcing}. 

In many of these studies adherence is taken to be synonymous with trust.  Indeed, while there is no single commonly adopted definition of trust in the HCI literature, the term trust typically refers to a measure of, or the factor influencing, the degree to which the human is willing to delegate decision-making to the machine in absence of complete knowledge of the algorithmic pipeline~\cite{lee2004trust, yu2016trust, bansal2019updates, nourani2019effects, yin2019understanding}.  We note that in our setting---and in any setting where the objective optimized for by the algorithmic system does not fully capture all relevant costs and payoffs of the decision---trust and adherence are not one and the same.   In the child welfare context, the risk assessed by the tool is just one factor relevant to determining whether an investigation is appropriate. Other factors such as resources, recent investigations, information conveyed during the allegation call, are all also relevant. Workers are not expected to adhere to the scores in all cases, and may trust the ability of the tool to assess certain dimensions of risk even if in a given instance that risk is not the most relevant decision factor.  Both ~\cite{chouldechova2018case} and ~\cite{brown2019toward} provide more indepth discussion of algorithmic tools and their use in the child welfare system.   Most notably, ~\cite{brown2019toward} take a deep dive into factors that influence affected communities perceptions comfort and trust in the use algorithmic decision support in the child welfare context.  The authors find that general distrust in the child welfare system, as well as specifics of how and whether risk information is communicated to case workers and families are important to perceptions of procedural and interpersonal justice.

% More broadly, while there is no single commonly adopted definition of trust in the HCI literature, the term trust typically refers to a measure of, or the factor influencing, the degree to which the human is willing to delegate decision-making to the machine in absence of complete knowledge of the algorithmic pipeline~\cite{lee2004trust, yu2016trust, bansal2019updates, nourani2019effects, yin2019understanding}. In the context of our analysis, it is possible to think of trust as a measure of how often and to which extent human's decisions are affected by the algorithmic tool. We have shown that the decisions of call workers are driven but not dictated by the machine's recommendations. This selective trust is a product of the workers not blindly relying on the machine, but instead posing its recommendations under continuous scrutiny.

Our work contributes to the nascent literature on human-machine complementarity in risk assessment by investigating whether humans are able to correct for misestimated risk scores in a real world decision making context. Our study relies on a retrospective analysis of observed call worker decisions before and after a risk assessment tool was integrated into the decision making pipeline. A similar context but in a different setting is analysed by~\cite{bushway2012sentencing}, which study the effects of inconsistencies in sentencing recommendations due to human errors in the judicial setting in Maryland. 
The authors find that judges are more likely to go along with mistakenly lesser sentences for violent offenses, but tend to discount recommendations that are mistakenly too high.  % such misestimation had a direct impact on judges' decisions. 

%The observational nature of our data has both advantages and disadvantages. One significant disadvantage is that we are unable to experimentally manipulate elements of the decision making setting to identify which factors are most relevant to driving observed outcomes. We do, however, discuss what we hypothesize to be some of the key contributing factors. % A clear advantage of our study design is that we are able to observe human-machine interaction in a real world setting.  It would be of considerable interest to devise controlled experiments to further 

\section{Deployment Setup and Data}
\label{sec:data}

Call workers at the child welfare hotline are tasked with deciding whether a call alleging potential child maltreatment or neglect should be screened in for investigation. In making their decisions, call workers in Allegheny County have access to the information communicated in the referral call, along with multi-system administrative data on demographics, child welfare involvement, criminal history, and other information related to the children and adults associated to a referral. The administrative data consists of hundreds of data elements.  It is therefore challenging for workers to make systematic and effective use of the administrative data in each case.  In order to help workers make better use of this data, Allegheny County introduced a risk assessment tool that distills the information contained therein into a single risk score reflecting the likelihood that the children on the referral will experience adverse child welfare related outcomes in the months following the referral.   The intended use of the tool is to help workers identify high-risk cases in instances where the information communicated in the call may be insufficient, inconclusive, or otherwise incomplete in reflecting the immediate or long-arc risk of the children.  As we further discuss in Section~\ref{subsec:dep}, specific guidelines were created to strongly encourage screen-ins (investigations) for the highest scoring cases.   %Throughout the paper, we refer to this decision as 

%In particular, cases predicted to be in the top 15\% highes-risk bucket require all cases where these is sufficient information 

\subsection{Predictive model}
Throughout the paper, we refer to the case associated to a call as a \textit{referral}, each of which may have several \textit{referral records} associated to it, one for each children involved in the call. The deployed tool was trained with all referral records collected by Allegheny County between April 2010 and July 2014. Two distinct predictive models estimate the probability of \textit{out-of-home placement} and of \textit{future referral} for each child based on features that include demographics, past welfare interaction, public welfare, adult and juvenile criminal justice involvement, and behavioral health information available on all persons associated with each referral. \textit{Out-of-home placement} refers to whether the child is placed out of the home following an investigation, and a \textit{future referral} refers to a future call involving the child coming in to the hotline. The predicted probabilities are then converted into an integer score in the range from $1$ to $20$, corresponding to the ventiles. The score to be shown to the workers is calculated as the maximum score over both models, over all children involved in a referral. We denote this aggregated score as $S \in \{1,\dots,20\}$. A more detailed description of the model can be found on the County's website~\cite{afst}.%\cite{chouldechova2018case}. %The resulting model was deployed in August 2016. 

%We use data from January 2015 to July 2016 to analyze the behavior of call-workers before the deployment of the tool, and data from August 2016 until December 2017 to study the call-workers behavior after adoption. 

\subsection{Deployment}
\label{subsec:dep}

\begin{figure}[ht]
    \centering
    \includegraphics[scale=0.2]{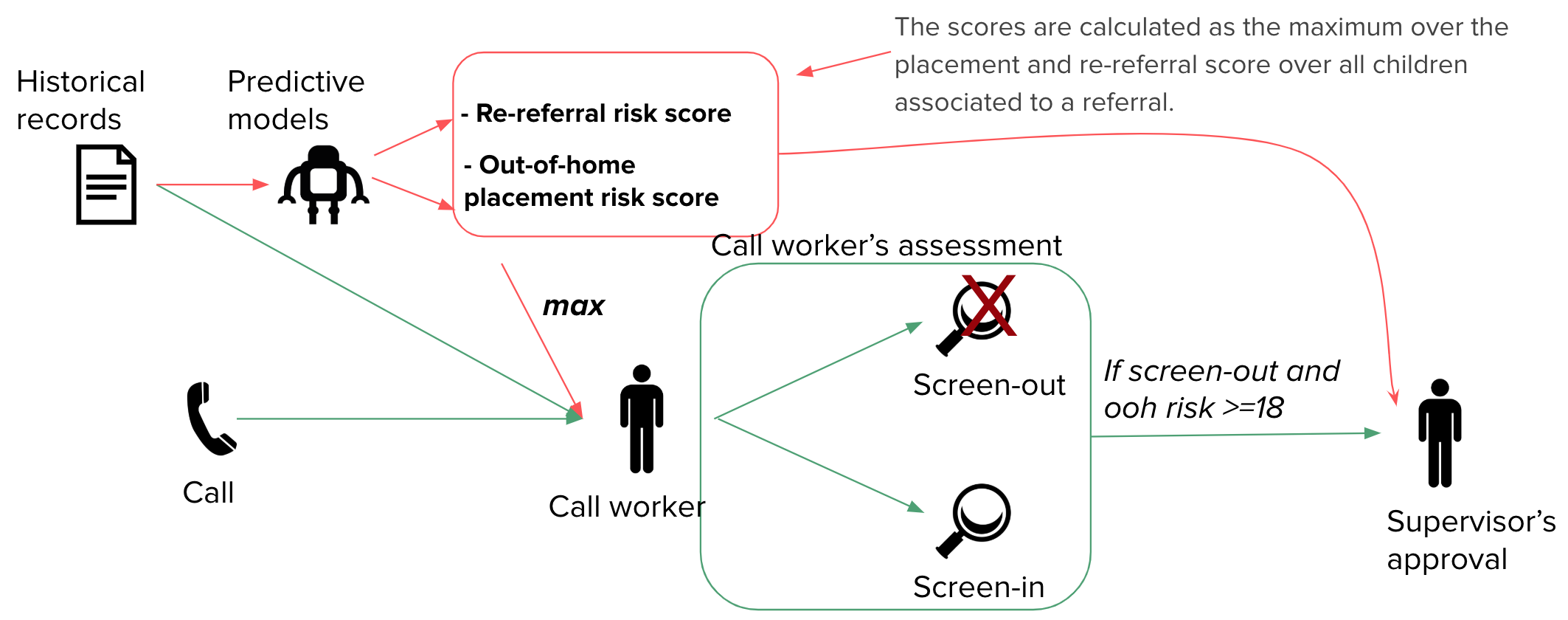}
    \caption{Decision pipeline for child welfare services in Allegheny County. Re-referral risk score and out-of-home placement (ooh) risk score are calculated for all children associated to a referral. The maximum over these scores is shown to the call worker, who also has access to historical records and information conveyed in the call. The call worker decides whether to screen in the call for investigation. Screening out calls with an ooh risk score $\geq$ 18 requires supervisor's approval. }
    \label{fig:decision_pipe}
\end{figure}

By design, call workers are only shown risk scores for cases with sufficient information. During the analyzed deployment period, $92.5\%$ of referrals had an associated score shown. In addition to displaying a risk score, the tool assigns a label of ``mandatory screen-in'' to certain referrals, which means that the supervisor's approval is required in order to screen out the referral. These correspond to the cases whose calculated score for out-of-home placement is greater or equal to 18. That is, these are cases where at least one child associated to the referral has a \textit{placement} score of at least 18. Figure ~\ref{fig:decision_pipe} presents a graphical representation of the decision-making pipeline.

\subsection*{Score misestimation}
\begin{figure*}[ht!]
    \centering
    \begin{subfigure}{.45\textwidth}
    \includegraphics[width=\linewidth]{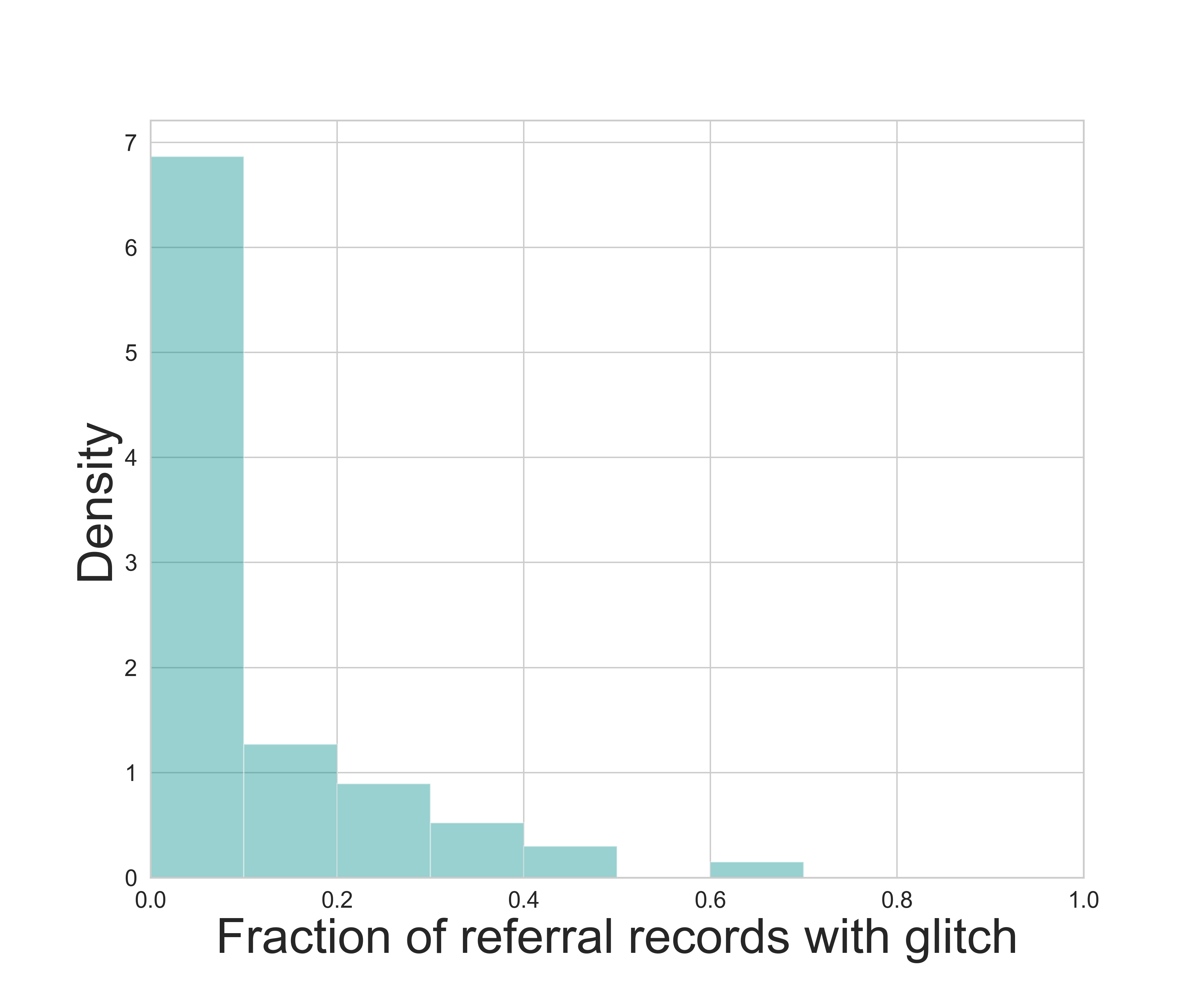}
    \caption{}
    \label{fig:glitch_barplot_by_feature}
    \end{subfigure}
    \begin{subfigure}{.45\textwidth}
    \includegraphics[width=\linewidth]{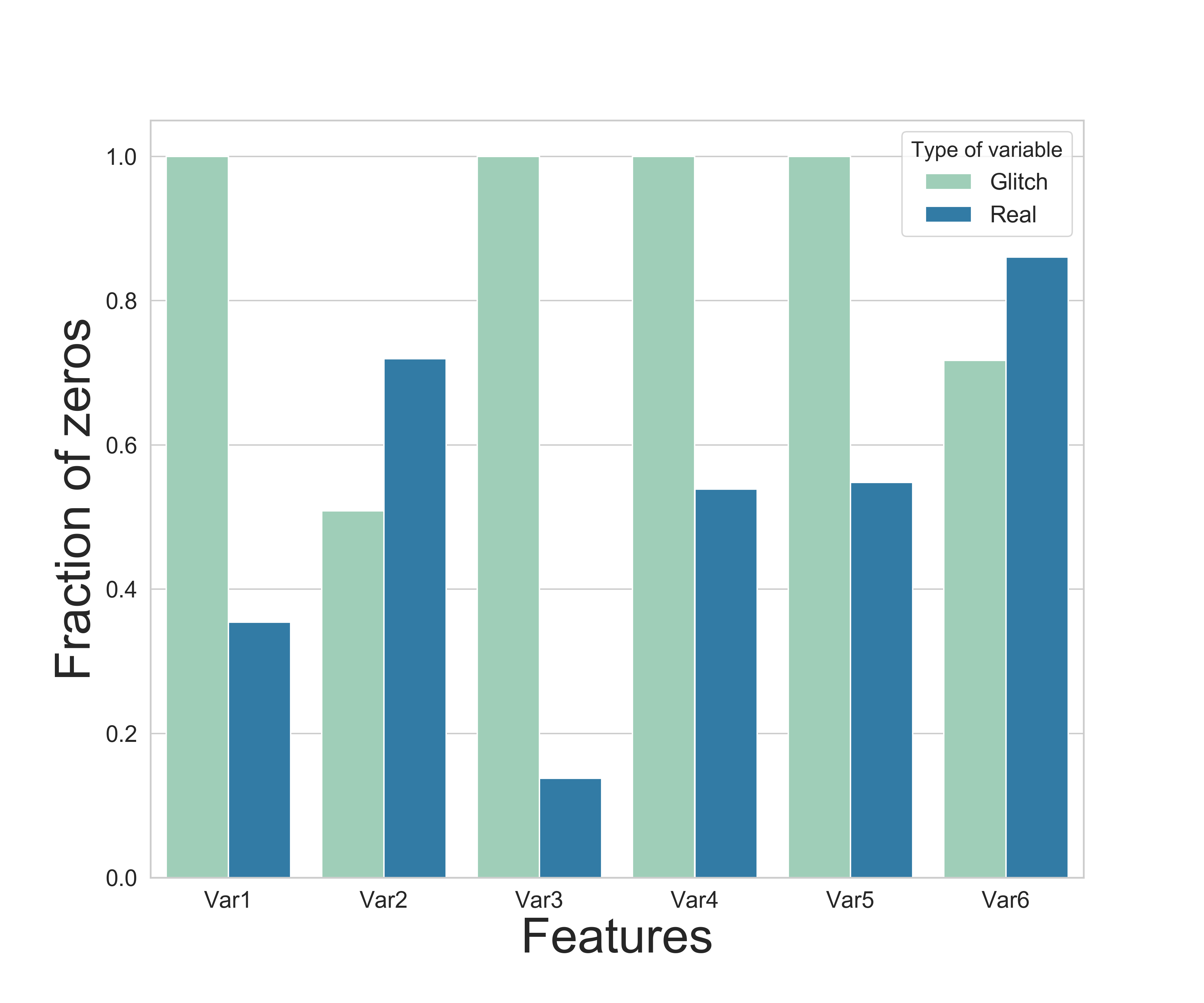}
    \caption{}
    \label{fig:glitch_hist_by_feature}
    \end{subfigure}
    \caption{Analysis of the glitch. (a) shows the normalized histogram of the fraction of referral records affected by the glitch for each feature. For the six (binary) features with highest fractions, (b) shows their distribution with and without the glitch.} 
    \label{fig:glitch_variables}
\end{figure*}

During deployment, a glitch in the system led to certain model inputs not being calculated correctly in realtime.  Figure~\ref{fig:glitch_barplot_by_feature} shows a histogram of the fraction of values for each feature whose realtime values did not match retrospectively re-calculated values. There are two reasons for this mismatch.  The first is due to an issue, since resolved, wherein the realtime database queries were erroneously returning counts and indicators of $0$. Figure~\ref{fig:glitch_hist_by_feature} compares the fraction of referral records where the feature was calculated as $0$ in real time vs. retrospectively for the top six most frequently mismatched binary indicators. The second issue is that as cases evolve the roles of different adults associated to the case and information about them may evolve. For instance, the individual identified as a perpetrator may change between the initial run and the retrospective analysis. These more naturally occurring mismatches are unlikely to have as significant an impact on the calculated score.  

\begin{figure}[h]
    \centering
    %\begin{subfigure}{.45\textwidth}
    \includegraphics[width=\linewidth]{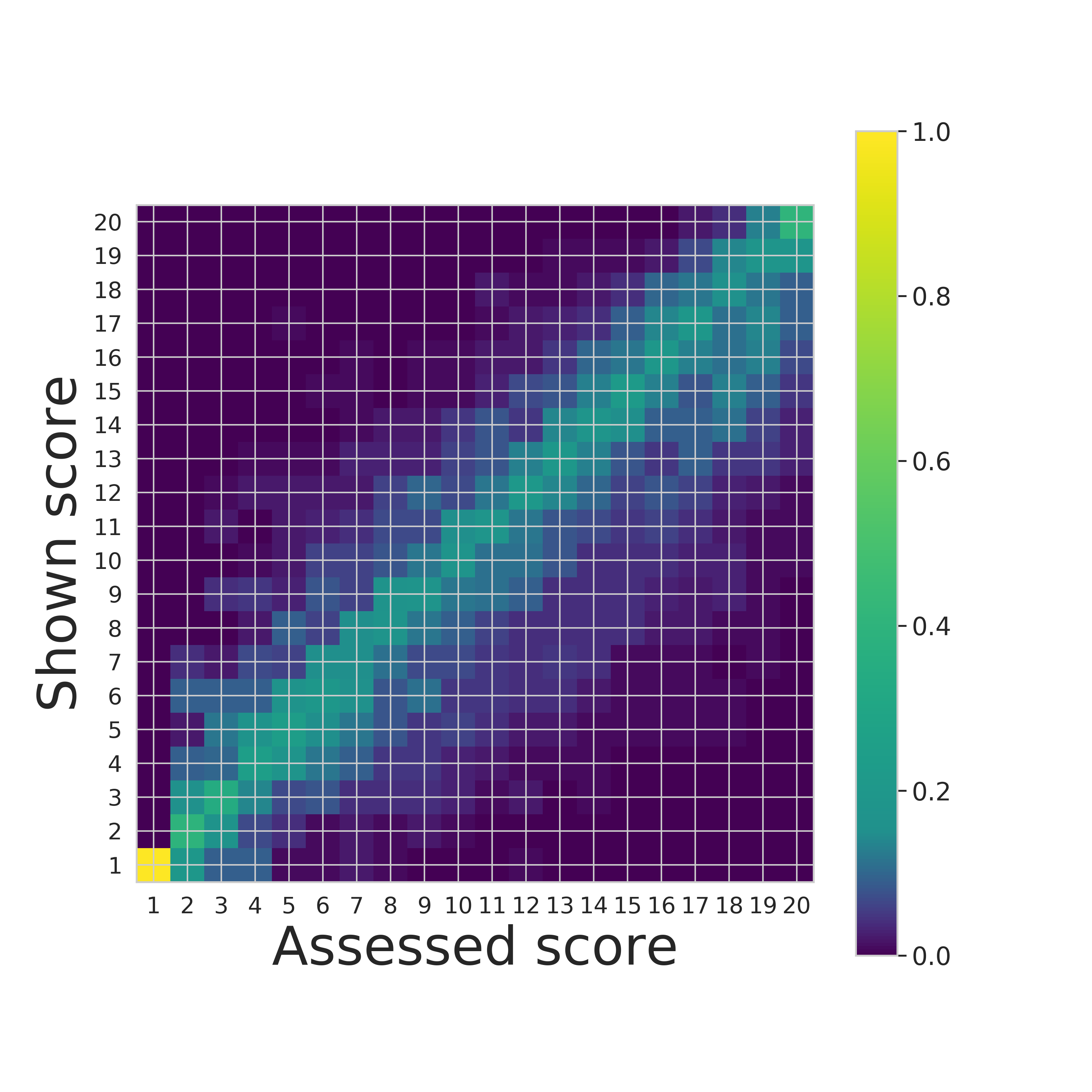}
    %\caption{ }
    % \label{fig:my_label}
    % \end{subfigure}
    % \begin{subfigure}{.53\textwidth}
    % \includegraphics[width=\linewidth]{barplot_scores.png}
    % \caption{ }
    % \end{subfigure}
    \caption{Heatmap of the density of shown score $\tilde{S}$ conditional on assessed score $S$.  A cell at row $r$ and column $c$ shows the fraction of the time that a referral assessed at a score of $S= c$ was shown to have a score of $\tilde S = r$.}%(b) Distribution of shown scores for each real score value. Boxes indicate median and quartiles, whiskers indicate 5-95 percentiles, and flier points are those past end of whiskers.   }
    \label{fig:scores_corrup}
\end{figure}

As a result of the glitch in how model inputs were being calculated in realtime, the score displayed (`shown') to workers during deployment did not always correspond to the score that should have been shown. Figure~\ref{fig:scores_corrup} shows the distribution of the scores. The concentration of cases around the diagonal indicates that the shown scores were most often equal or very close to the assessed score that should have been displayed. These circumstances of deployment allow us to study the behavior of call workers when the shown score is inaccurate. In particular, we are able to analyze what happens when differences between the assessed and the shown score result in a different ``mandatory screen-in'' policy being applied.

The terminology and notation we use throughout the paper are detailed in Table~\ref{table:notation}. 

\begin{table*}[ht!]
\caption{Terminology and notation used throughout the paper.}
\label{table:notation}
\begin{tabularx}{\linewidth}{p{0.22\linewidth}p{0.12\linewidth}p{0.59\linewidth}}
\toprule
\textbf{Term}   & \textbf{Notation}      & \textbf{Description } \\ \midrule
\textit{Assessed score} & $S \in \{1,\dots,20\}$  & Score assessed by the predictive model. Maximum over assessed risk of out-of-home placement and assessed risk of re-referral.         \\ \hline
\textit{Shown score}  &$\tilde{S} \in \{1,\dots,20\}$  & Score shown to the call worker. It should correspond to the assessed score, but is subject to a glitch in the system.        \\ \hline
\textit{Assessed mandatory screen-in}  & $M \in \{0,1\}$  &  $M=1$ if assessed risk of out-of-home placement greater or equal to 18. Supervisor's approval required for screen out.      \\ \hline
\textit{Shown mandatory screen-in } & $\tilde{M} \in \{0,1\}$ &  Mandatory screen-in label shown to the call worker.  It should correspond to \textit{assessed mandatory screen-in}, but is subject to a glitch in the system.          %  \\\hline
%\textit{Screen-in decision } & $D\in\{0,1\}$ &   Call workers' decision to screen in the call. 
\\ 
\bottomrule
\end{tabularx}
\end{table*}

%Throughout the paper, we refer to the predicted score as the \textit{real score}, and to the (sometimes miscalculated) score displayed by the tool as the \textit{shown score}, and denote them by $S, \tilde{S} \in \{1,\dots,20\}$, respectively; $M, \tilde{M} \in \{0,1\}$ are the predicted and shown mandatory screen-in labels, respectively; $D\in\{0,1\}$ is the human's decision to screen in a call. %\fixm{Update all ``real score'' to ``predicted score''? Real may be interpreted as a reference to some ground truth so it may bother some readers.}

%We refer to this label as $M \in \{0,1\}$. 

%As such, a ``mandatory screen-in'' policy dictates that cases with a predicted risk of out-of-home placement greater or equal to 18--which corresponds to the $15\%$ riskier cases--, which require a hotline supervisor to override the decision if the call-worker assesses the call should not be screened in.

\subsection{Data}

The risk assessment tool was deployed in August 2016. We use data from January 2015 to July 2016 to analyze the behavior of call workers before the deployment of the tool, and data from August 2016 until December 2017 to study their behavior after adoption. 

We constrain our analysis to the $92.5\%$ of referrals that had an associated score shown. Moreover, existing regulation dictates that certain calls must be investigated. Referrals that fall under this category include those in which the call concerns bodily injury and sexual abuse. Since the call worker had no discretion on deciding whether or not to screen in these calls, we exclude from our analysis the 19\% of referrals that fall under this legislation. Finally, we also exclude the referrals that are associated to open investigations, since the call worker does not make any determination in these cases. This excludes 19\% of the remaining observations. Once we have constrained the data to those cases that had a score shown and for which the call worker made a determination, we are left with 27,575 referral records, corresponding to 11,802 referrals (recall that multiple children may be associated to a call, leading to multiple referral records within each referral).% \fixm{Update this number.} %maybe add the details of what info we have associated to each referral?
% For the exclusion based on the call_scrn, I changed 15% to 19% because we are also excluding the referrals with unknown/pending call screen code

\section{Analysis}
\label{sec:analysis}

\subsection{Change in call workers' behavior}
\label{sec:adherence}
% Stanford evaluation: would their results change now that they know that the scores are corrupted?

%Although the risk assessment tool was deployed only in August 2016, retro-actively computed scores for the period August 2014 to July 2016 are available. 

The main question we seek to answer in this section is whether call workers updated their screening behavior after the tool was deployed. In other words, did algorithm aversion lead to the tool being ignored altogether?  %To answer this, we do two types of analysis. First, we analyze how the relationship between screen-in rates and predicted scores changes pre- and post-deployment. The second part of the analysis does not rely on scores, and instead looks at the predictability of human decisions.   

We first analyze the behavior of call workers with respect to the assessed score, $S$, before and after the deployment of the tool.  While ideally we would also display similar results for the shown score, $\tilde{S}$, the glitch that produced $\tilde{S}$ is not retrospectively reproducible.  We are unable to calculate what $\tilde S$ would have been in the pre-deployment period.  %  While humans observe $\tilde{S}$ instead of $S$, we do our analysis this way because it is not possible to calculate what the shown score, $\tilde{S}$, would have been in the pre-deployment period. 
However, since pre- and post-deployment decisions would remain the same with respect to both $S$ and $\tilde{S}$ if workers did not alter their decisions, it is nevertheless insightful to study variation with $S$.  
%Given the tight relationship between $S$ and $\tilde{S}$ shown in Subsection ~\ref{subsec:dep}, and given the fact that if humans did not update their behavior, then the relationship pre- and post-deployment should remain the same with respect to both $S$ and $\tilde{S}$, this setup allows us to answer our question of interest.  %this analysis is likely to be a conservative estimate of humans' alignment with the predicted score $S$ post-deployment. W

%Our analysis in the current section investigates {\it if } and {\it how} the deployment of the tool has affected the behavior of the call-workers. 

\begin{figure}[t]
    \includegraphics[width=\linewidth]{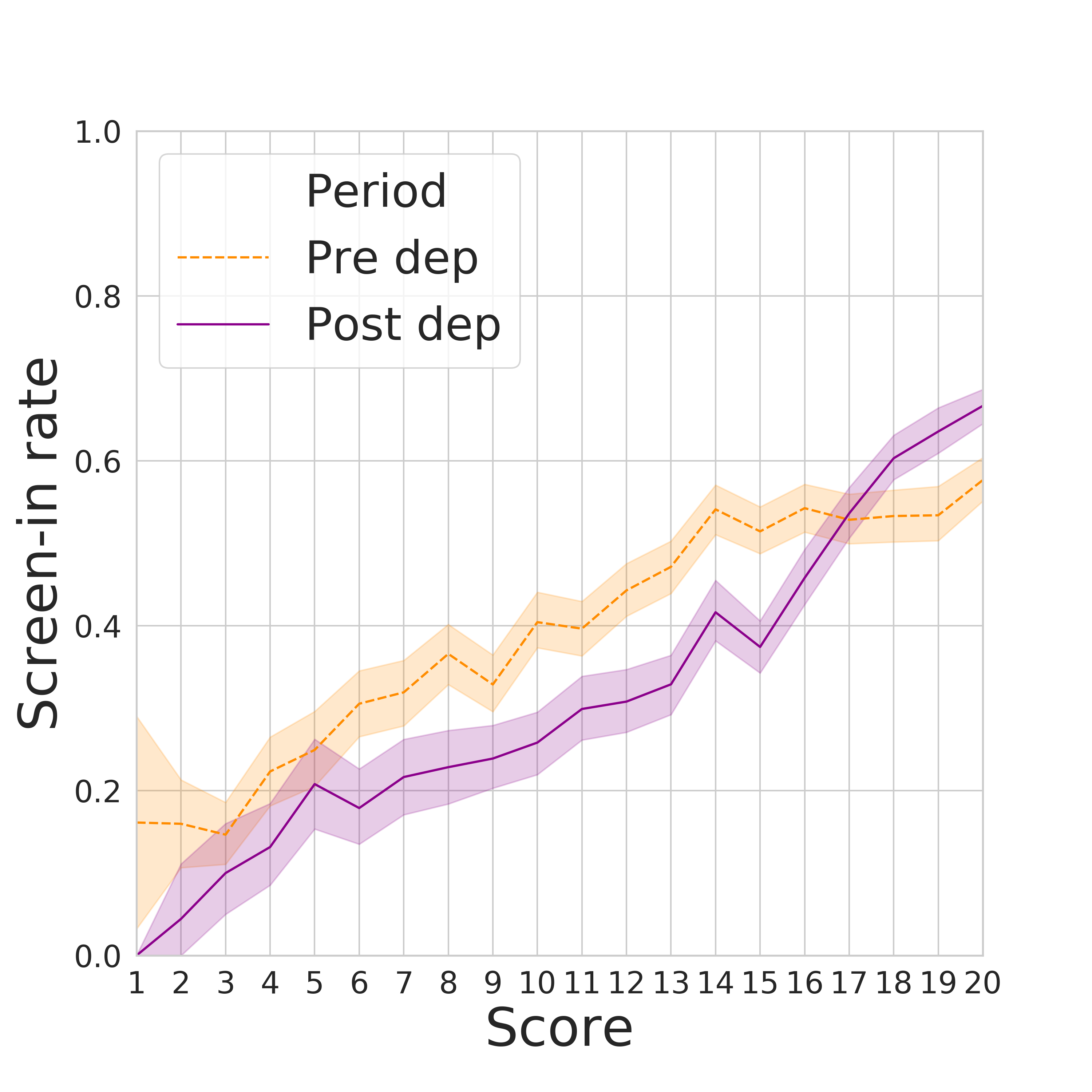}
    \caption{Screen-in rates by assessed score $S$ before and after deployment. After deployment screen-in decisions are better aligned with the score. }
    \label{fig:pre_post}
\end{figure}

As we now discuss, the results indicate that workers did update their behavior. First, we note that the overall screen-in rate did not vary before and after deployment, remaining around $45\%$. This stability over time is likely explained by the fact that the agency investigates as many cases as their resources allow. However, we do observe a change on {\it which} cases were investigated. Figure~\ref{fig:pre_post} shows the screen-in rates across values of the assessed risk score $S$. The steeper slope of the post-deployment curve, particularly for very high and very low risk cases, indicates that post-deployment screen-ins are better aligned with the score.  We see a pronounced increase in the screen-in rate of highest risk cases, and a pronounced decrease in the screen in rates for low and moderate risk cases.  
%In particular, we observe a trade-off between increased screen-in's for the highest risk cases and decreased screen-in's for the other scores, in particular for mid-range cases. 

\begin{figure}[t]
 \includegraphics[width=\linewidth]{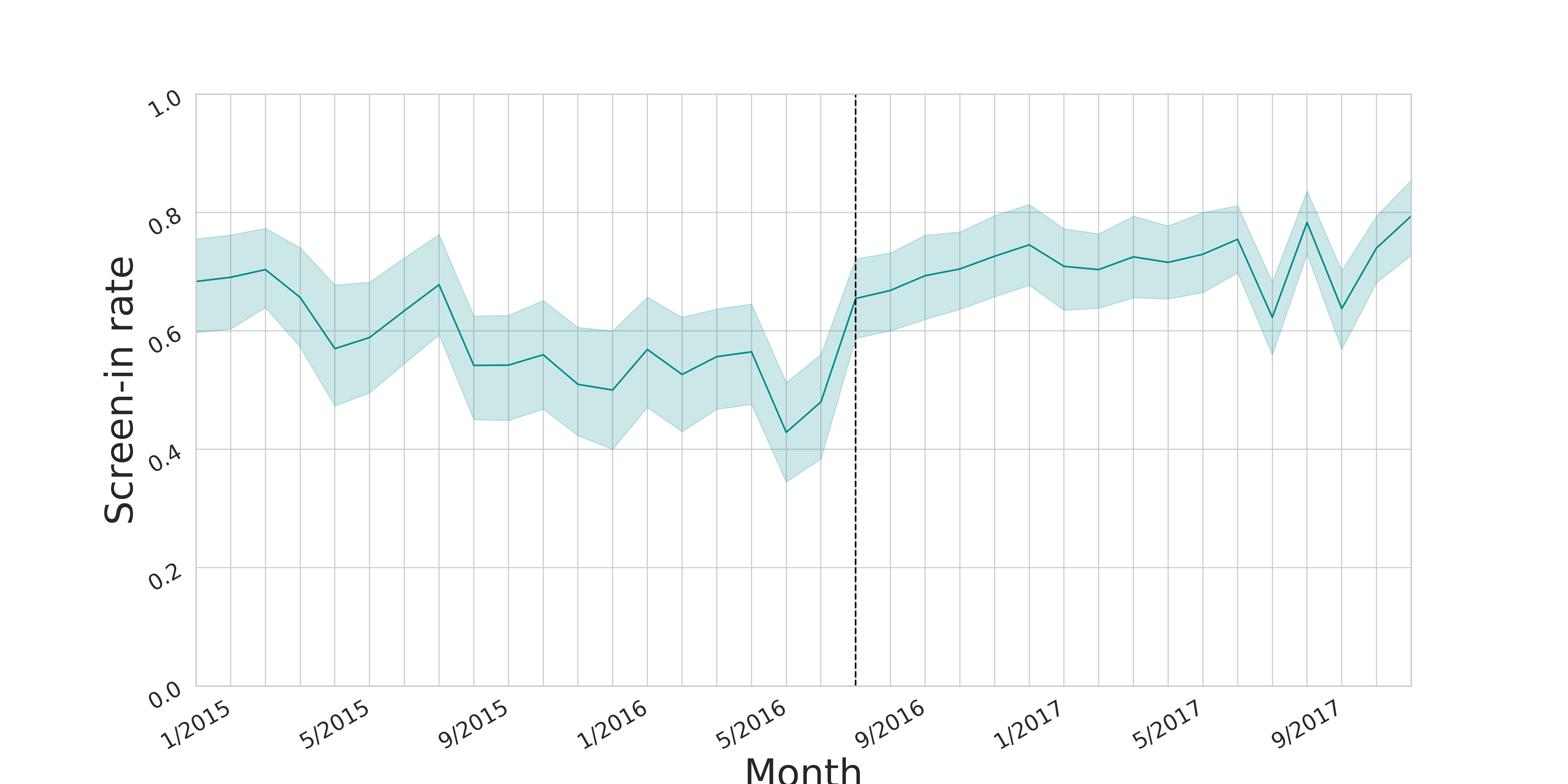}
 \caption{  Timeline of screen-in rates for cases with mandatory screen-in policy, $M=1$. Dotted line marks date of deployment.  We find a sharp and persistent increase in screen in rates for assessed mandatory cases.  Some of this increase is attributable to assessed mandatories that were also correctly shown as mandatory screen-ins. }
    \label{fig:timeline_PL}
\end{figure}

Recall that any referral where the placement score was 18 or greater was flagged as a ``mandatory screen-in''.  Shown mandatories required supervisory approval to be screened out.  While not all assessed mandatories ($M = 1$) where shown mandatories ($\tilde M = 1$) and vice versa, it is nevertheless instructive to look at how screen-in rates varied pre- and post- deployment with respect to $M$.  
 Figure~\ref{fig:timeline_PL} shows a timeline of screen-in rate of cases corresponding to $M=1$. A sharp increase in screen-in rates for assessed mandatory cases can be seen the month of deployment. Overall, in the period before deployment, cases that would have fallen in the category $M=1$ had a screen-in rate of $58\%$, while after deployment this rate increased to $71\%$. As there appears to be no change in the jurisdiction's policies or types of calls received around the time of deployment, we attribute the sudden increase in screen-in rate to the deployment of the model. This increase is to be expected given that there is a barrier to overriding cases $\tilde{M}=1$, which is strongly correlated with $M$.  Notably, the increase is smaller than what would be expected if automation bias were to occur.  Even though there is a cost-barrier to override the machine (required supervisory approval), only $66\%$ of cases shown as mandatory screen-in, $\tilde{M}=1$, are screened in. %The relationship between assessed and shown scores and labels is explored more in detail in Section~\ref{sec:adherence}.% This shows that while call workers still seek supervisor's approval to screen out many of these cases, the deployment setup did lead to an increase in investigations for these cases.      

\subsection{Overrides of erroneous scores}
\label{sec:correction}

\begin{figure}[t!]
    \centering
    \includegraphics[width=\linewidth]{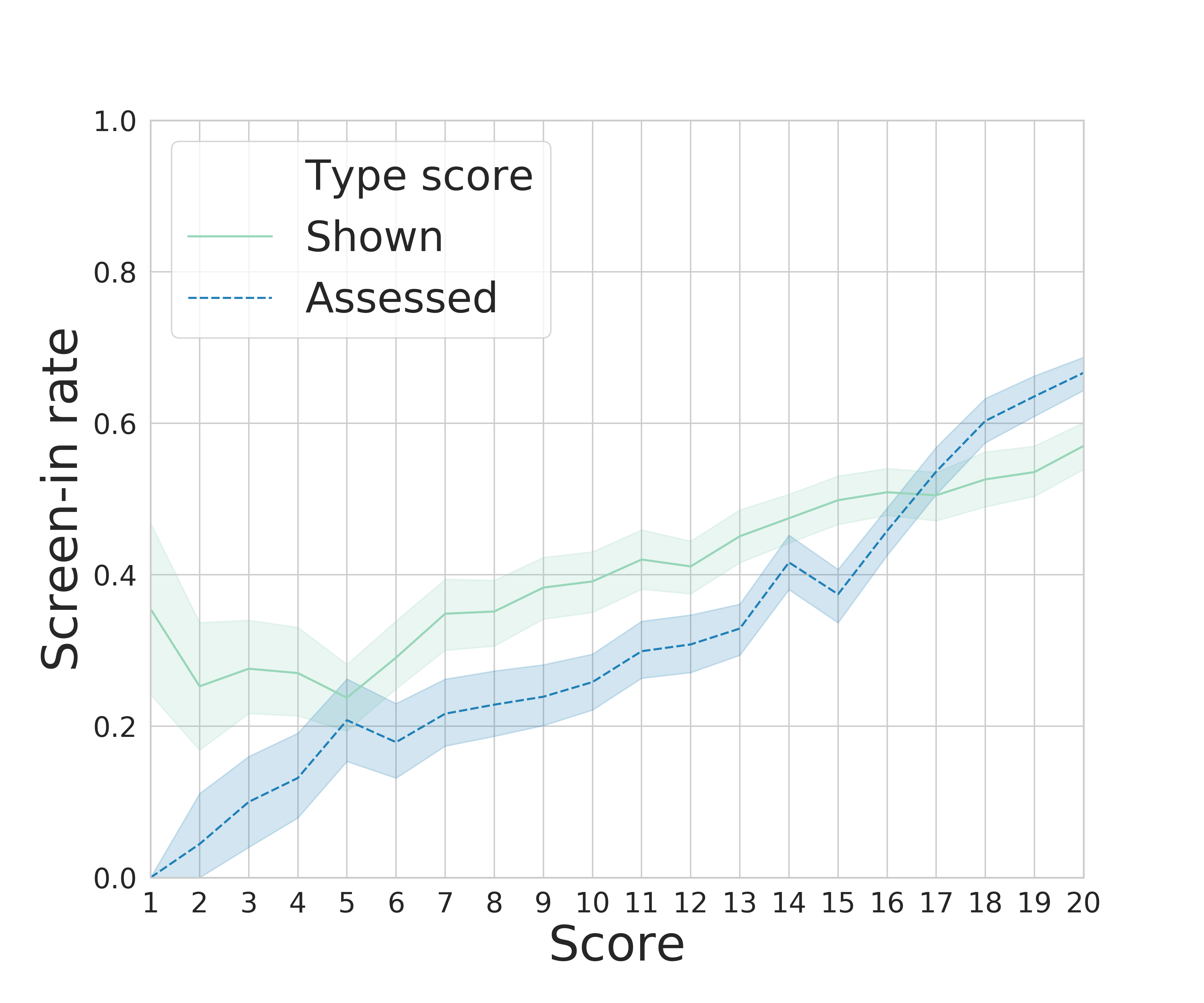}
    \caption{Alignment of screen-in rates with respect to assessed score $S$ and shown score $\tilde{S}$. We find that post-deployment decisions are better aligned with the assessed score than the score that was shown, indicating that workers were successful in correcting for the score miscalculation in certain cases.}
    \label{fig:calib}
\end{figure}

%(Include new subsection) Is performance better with the tool?--include this on 4.1 or 4.2, respectively-> Percentage of cases with services offered. In this new section focus on disparities. black-white disparities and disparities across poverty ranges are reduced, impact on age only for very young kids, for whom screen-in rates increase. 

\begin{figure*}[t!]
    \centering
    \begin{subfigure}{.45\textwidth}
    \includegraphics[width=\linewidth]{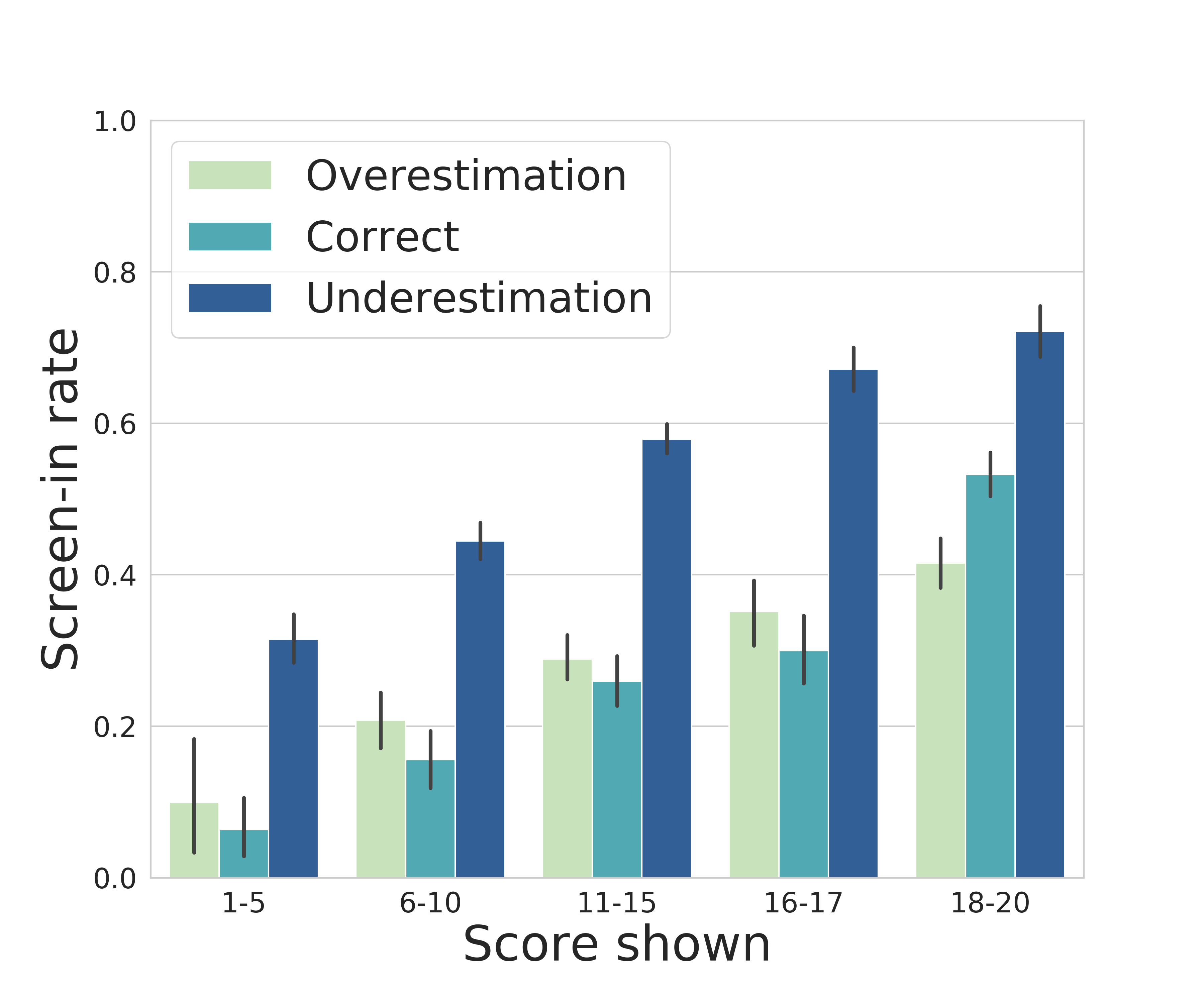}
    \caption{ }
    \label{fig:human_correction_a}
    \end{subfigure}
    \begin{subfigure}{.45\textwidth}
    \includegraphics[width=\linewidth]{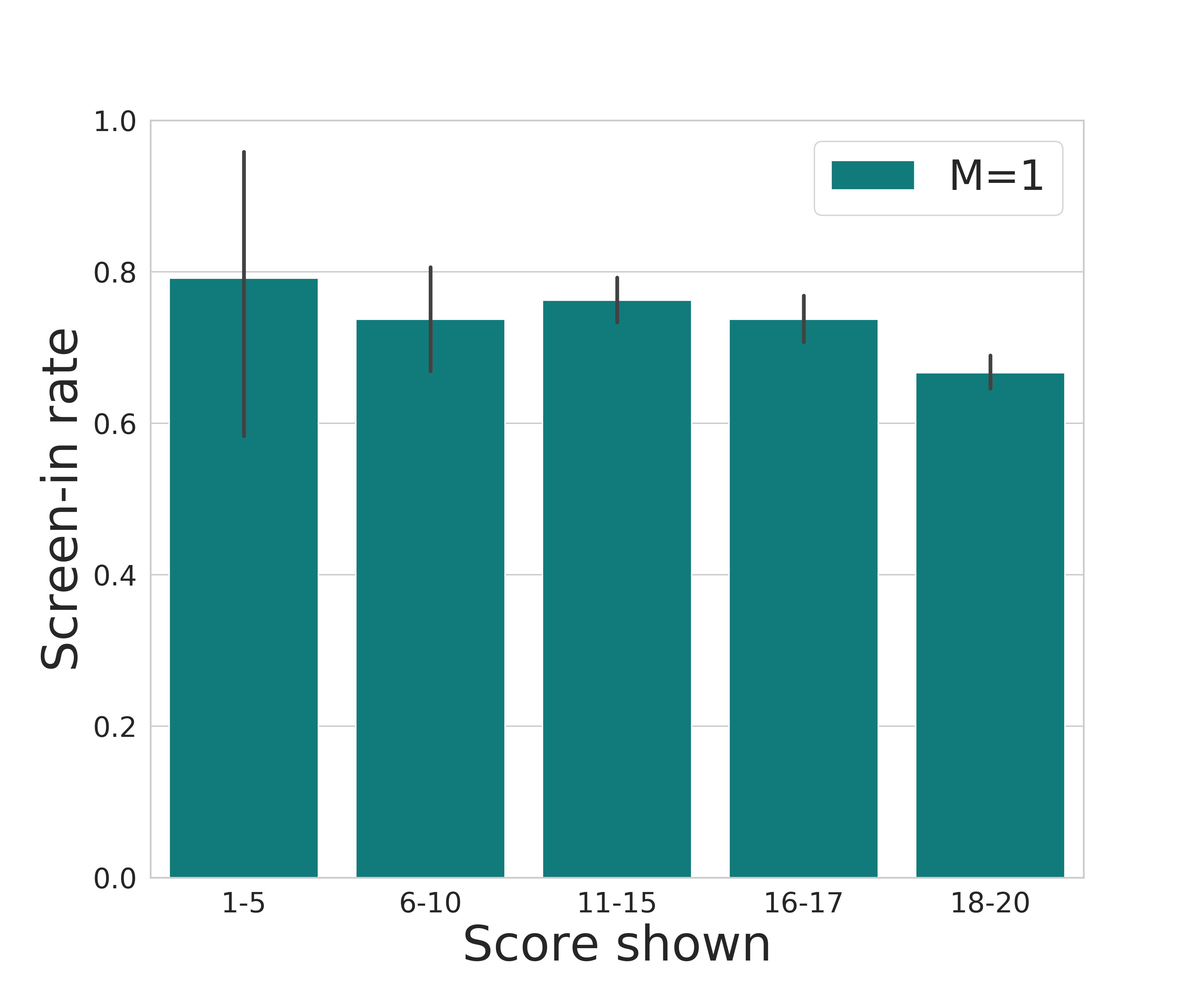}
    \caption{ }
    \label{fig:human_correction_b}
    \end{subfigure}
    \caption{Analysis of human decisions. Fraction of screen-in's for: (a) overestimation ($\tilde{S}> S$), equality ($\tilde{S}=S$), underestimation ($\tilde{S}< S$) of the assessed score for binned shown scores; (b) binned shown score for cases in which the screen-in should have been mandatory, $M=1$. Error bars indicate $95\%$ confidence intervals.} %computed via bootstrap.}
    \label{fig:human_correction}
\end{figure*}

In this section we investigate whether call workers detect and correct mistakenly computed algorithmic scores.  We study both types of errors common to the phenomenon of automation bias: omission errors and commission errors.  That is, we investigate whether call workers commit errors of omission---screening out shown low-risk cases that are assessed as high(er) risk--- and errors of commission---screening in shown high-risk cases that are assessed as lower risk.  % That is, do call workers appear to be suscetiple to automation bias, either through omission errors wherein they follow the recommendations when they are likely wrong, or commission errors wherein they override the recommendations when the recommendations are reasonable. % ? Or, do commission errors lead humans to override the machine at the wrong times, failing to make use of the information they have access to in order to successfully correct the machine's mistakes?
For this analysis, we use the fact that the shown score did not always correspond to the predicted score during the analyzed post-deployment period. Therefore, in some cases the score shown corresponded to an underestimation of risk, while in others it corresponded to an overestimation.

It is remarkable to note that the human decisions are better calibrated with respect to the assessed score $S$ than with respect to the shown score $\tilde{S}$, as seen in Figure~\ref{fig:calib}. The relatively high screen-in rates for cases with very low shown scores suggest that the call workers are effectively using information at their disposal to avoid many errors of omission, and choosing to screen in around $30\%$ of cases that are shown to have the lowest risk scores. Meanwhile, only a very small fraction of cases with an assessed score in the lowest buckets were screened in. This is consistent with what we observe at the high risk end of the scales. Again, call workers seem to be making use of available information to choose not to screen in cases that the machine marks as high-risk, thereby avoiding errors of commission. This result is particularly surprising given that overriding many of these cases would require a supervisor's approval. When focusing on those that required approval to be screened out, $\tilde{M}=1$, call workers screen in $66\%$ of the cases. This rate goes down to $57\%$ when only considering those that should not have been in this bucket (cases cases where $\tilde{M}=1$ and $M=0$). This indicates that humans are more likely to override a shown mandatory screen in when it is not an assessed mandatory.  It is not surprising that screen in rates are high for shown mandatories overall, as high shown scores correlated strongly with high assessed scores.  %(because shown and assessed scores are generally close to each other, most of these are likely to be close to 18, which explains why the screen-in rate is still higher than average). 

In order to explore this behavior further, we look at decisions for under- and over-estimated scores in Figure~\ref{fig:human_correction}. Figure \ref{fig:human_correction_a} shows the percentage of screen-in's for binned $\tilde{S}$. \textit{Correct} indicates that the assessed score is equal to the shown score, \textit{underestimation} means that the score shown was \textit{lower} than the assessed score, and \textit{overestimation} means that the score shown was \textit{higher} than the assessed score. If the call workers were to blindly follow the risk tool, we would observe that within each score bucket the screen-in rates are the same across all three classes. We observe something very different. It is clearly evident from the analysis that, among cases with similar shown scores, cases for which the score was underestimated are screened in at \textit{much} higher rates than the others. What this means is that humans are seemingly able to identify that the risk is being underestimated for these cases. For example, for the cases with a shown score between 11 and 15, those for which this score was an underestimation of the assessed score were screened in at a rate of almost 60\%, while the other cases with a shown score between 11 and 15 were screened in at rates around 30\%. This suggests that the call workers make use of other pieces of information---either from the call or from the administrative data system directly--- and respond appropriately by screening-in with higher probability. Somewhat surprisingly, we do not observe such a stark pattern for overestimated cases except in the highest risk bucket, where call workers are less likely to screen-in cases that should have had lower scores assigned.

While Figure \ref{fig:human_correction_a} allows us to observe certain patterns, it does not differentiate between over and underestimations of different magnitude. For that reason, we zoom-in on the decisions made for  assessed mandatories ($M=1$). Figure~\ref{fig:human_correction_b} shows the proportion of cases $M=1$ that were screened in for each bucket of shown score, $\tilde{S}$. This rate is approximately constant across all buckets, suggesting that the underestimation of the assessed score had no effect on the call workers' actions and they were able to make use of other information to identify these high risk cases. Even when the shown placement score was more than 12 points lower than the assessed placement score, the screen-in was still as likely as for a shown mandatory case. %The slight decrease in screen in rate as the score increases may be related to the nature of the glitch: scores where not randomly miscalculated--instead certain features were mis

%First, cases of underestimation are screened-in more frequently than cases of overestimation. Second, except for the highest bin where the mandatory screening plays a key role, the human seems to screen-in approximately the same proportion of cases with underestimation and with no corruption. This pattern could be a product of the determinism of the corruption mechanism: the human might care about a certain set of features that differ substantially between the corrupted and non corrupted cases. 

%Figure \ref{fig:human_correction} (b) shows the fraction of screen-in's separated by binned shown placement score for all cases for which screen-in would have been mandatory. 

%. The minimum size for the observations to be considered is 10

The first part of our analysis established that workers did update their behavior when the tool was deployed.   Meanwhile, the findings summarized in Figure~\ref{fig:calib} and Figure~\ref{fig:human_correction} indicate that they do not follow the score blindly. Instead, workers successfully make use of other sources of information and are more likely to override the machine when the shown score is significantly miscalculated. 

%One natural objection at this stage is that our discussion of the shown score's `correctness' hinges on a comparison to the assessed score.  

What can we say about the quality of these decisions, beyond their relationship with the assessed predicted risk? When a call is screened in, a social worker visits the family and is tasked with deciding whether to open a child welfare case, which is termed as ``accept for service". If call workers' performance had been perfect and there were no resource constraints, we could expect all referrals they screen in to have an case opened.  Before deployment of the tool, $18\%$ of investigated referrals were accepted for services, $19\%$ were connected to an existing open case involving the family, and $63\%$ were not accepted for service. Post-deployment, these rates were $21\%$, $23\%$, and $56\%$, respectively. This indicates a higher precision in the post-deployment period: more of the screened-in referrals were being provided with services. % a slightly higher precision of cases that indeed warrant an investigation, and lower false positive screen-in's. 

\begin{figure}
    \centering
    \includegraphics[width=\linewidth]{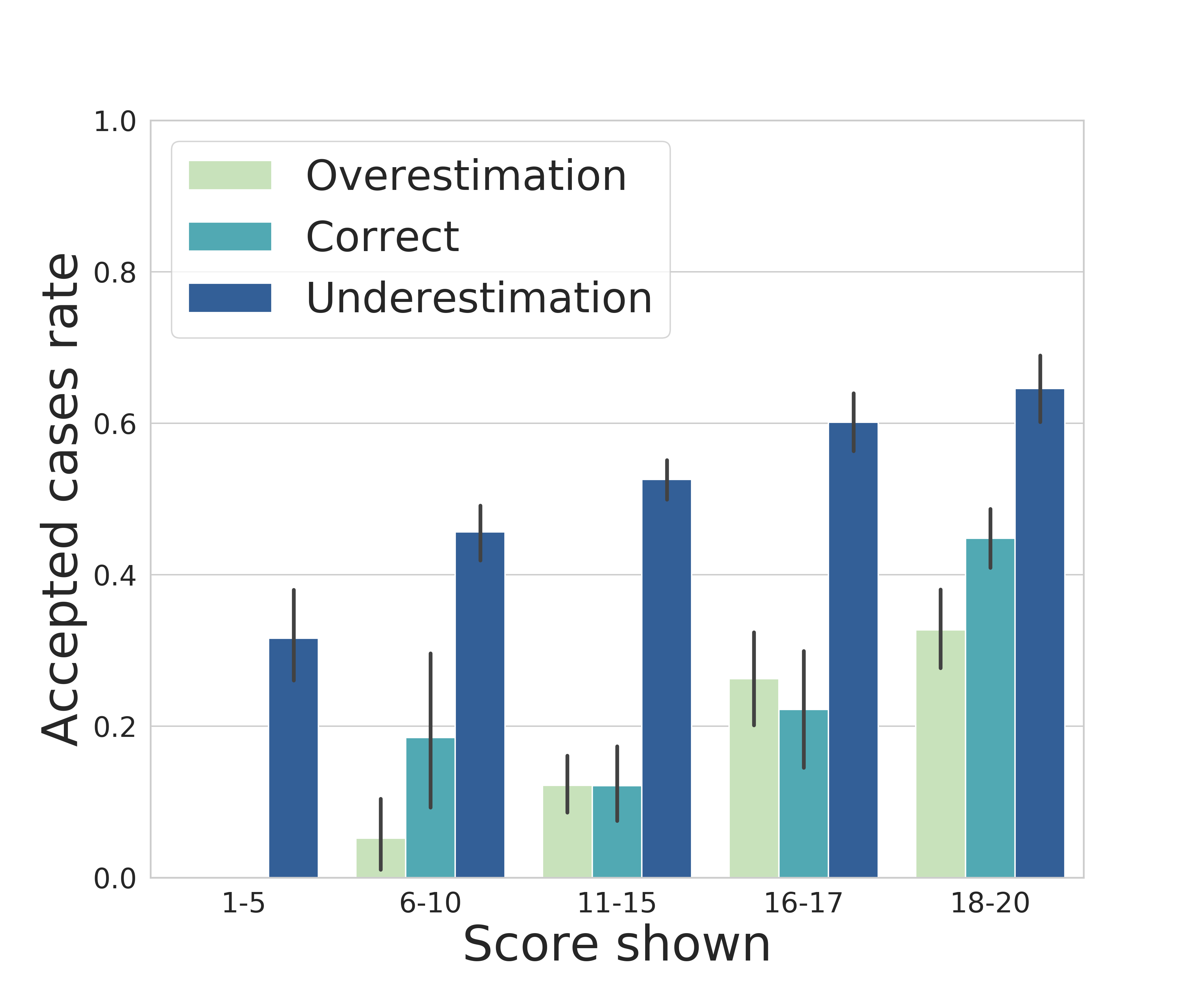}
    \caption{Rate of screened-in cases accepted for service across buckets of score shown, grouped according to the relationship between the score shown and the assessed score. }
    \label{fig:accept}
\end{figure}

Figure~\ref{fig:accept} shows the rates of cases accepted for service across buckets of scores, grouped according to the relationship between the score shown and the assessed score--the same grouping used in Figure~\ref{fig:human_correction_a}. For example, for the cases with a shown score between 11 and 15, more than half of those for which this score was an underestimation of the assessed score were accepted for services upon screen-in. Meanwhile, this statistic drops to 15\% for all other cases with a score shown in this bucket. The similarity between Figure~\ref{fig:accept} and Figure~\ref{fig:human_correction_a} is striking. In Figure~\ref{fig:human_correction_a} we observed that humans are very good at identifying cases that had an underestimated score, being more likely to screen these in. In Figure~\ref{fig:accept} we observe that these are indeed more likely to be accepted for service following an investigation.

\subsection{Disparities in decision-making}
\label{sec:disparities}

One of the main concerns surrounding the implementation of risk assessment tools is that this may exacerbate disparities, harming already marginalized groups. In the scenario of child welfare, disparities across race and income level are of particular concern \cite{eubanks2018automating}. While it is important to ensure that children are being protected, it is also important to avoid overburdening a group with interventions that may have harmful unintended consequences. 

\begin{figure}[t]
    \centering
    \begin{subfigure}{.45\linewidth}
    \includegraphics[width=\linewidth]{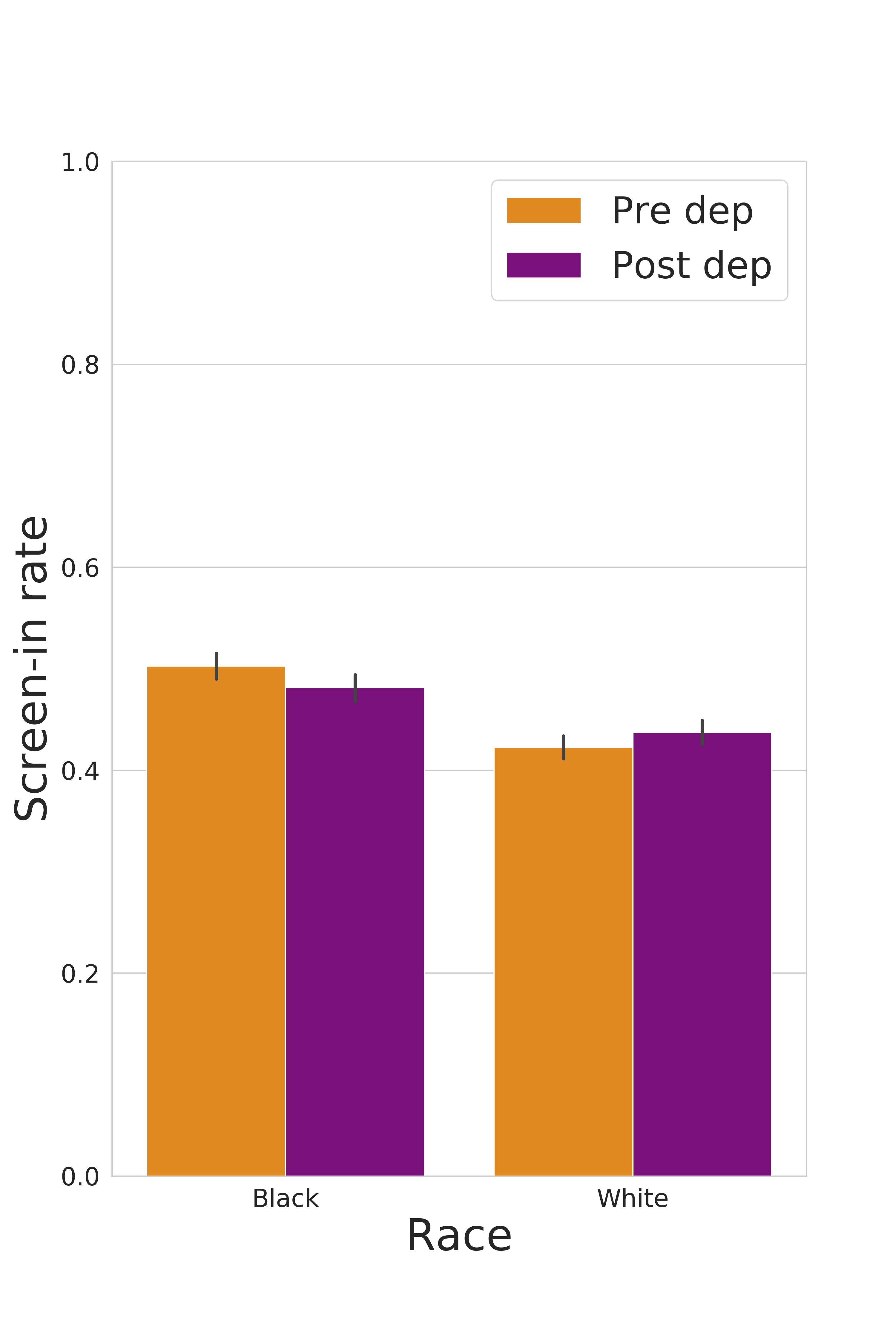}
    \caption{ }
    \label{fig:race_a}
    \end{subfigure}
    \begin{subfigure}{.45\linewidth}
    \includegraphics[width=\linewidth]{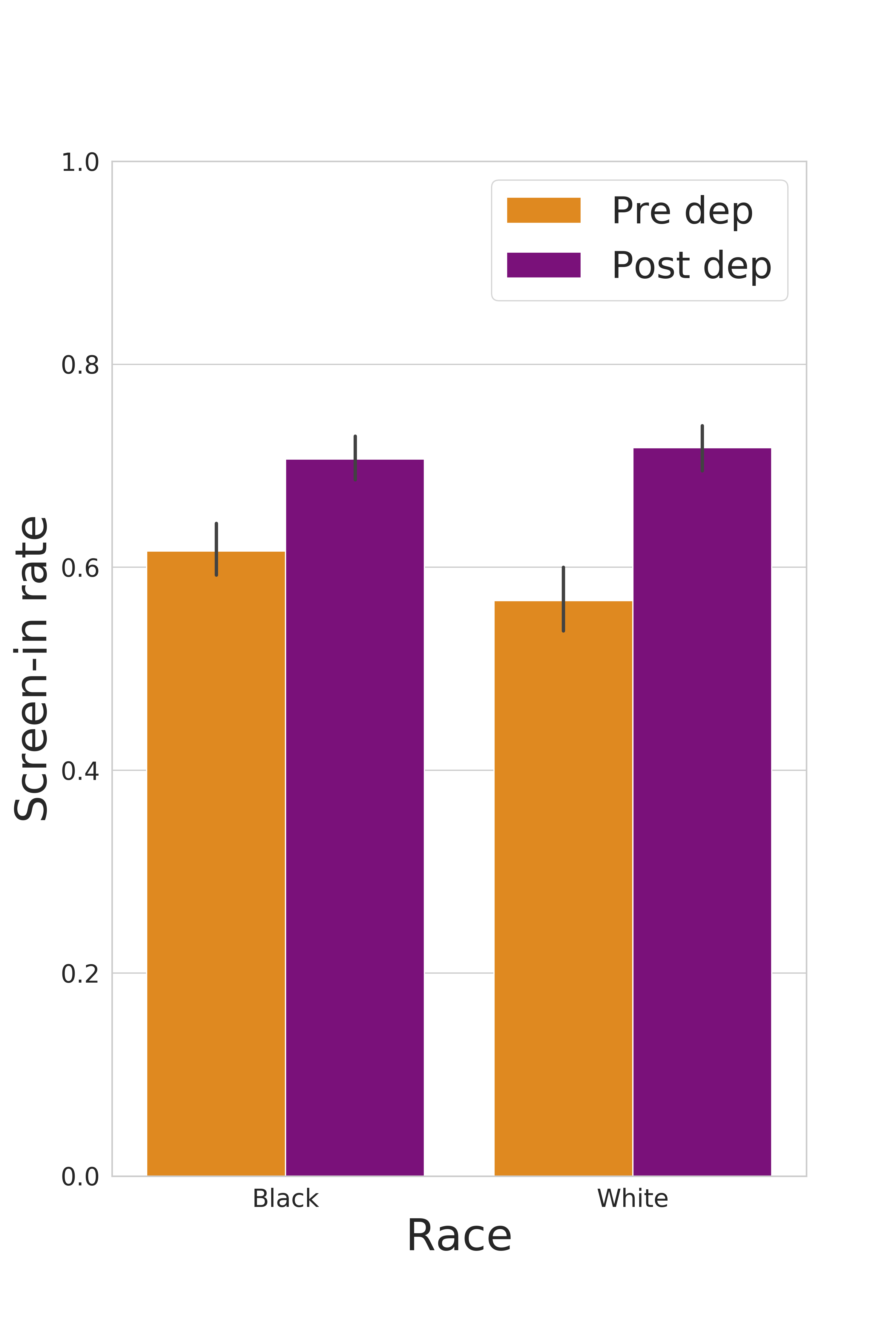}
    \caption{ }
    \label{fig:race_b}
    \end{subfigure}
    \caption{Screen-in rates by race pre- and post-deployment. (a) overall; (b) for subset of cases $M=1$.} %computed via bootstrap.}
    \label{fig:race}
\end{figure}

Figure~\ref{fig:race} shows the screen-in rates for Black and White children before and after the deployment of the tool. We observe a slight reduction on screen-in rates for Blacks and a slight increase for Whites. Prior work has identified differences in overrides across groups as a source of disparities. We therefore also analyze if there is a difference in alignment with the assessed mandatory screen-in policy across groups. Figure~\ref{fig:race_b} shows the screen-in rates for the subset of cases for which $M=1$. Here, we observe that screen-in rates for both races increase, although the increase is slightly sharper for White children. These results indicate that, unlike what has been observed in other domains, there does not appear to be a difference in willingness to adhere to the recommendation that would compound previous racial injustices.  

\begin{figure}[h]
    \centering
    \includegraphics[width=\linewidth]{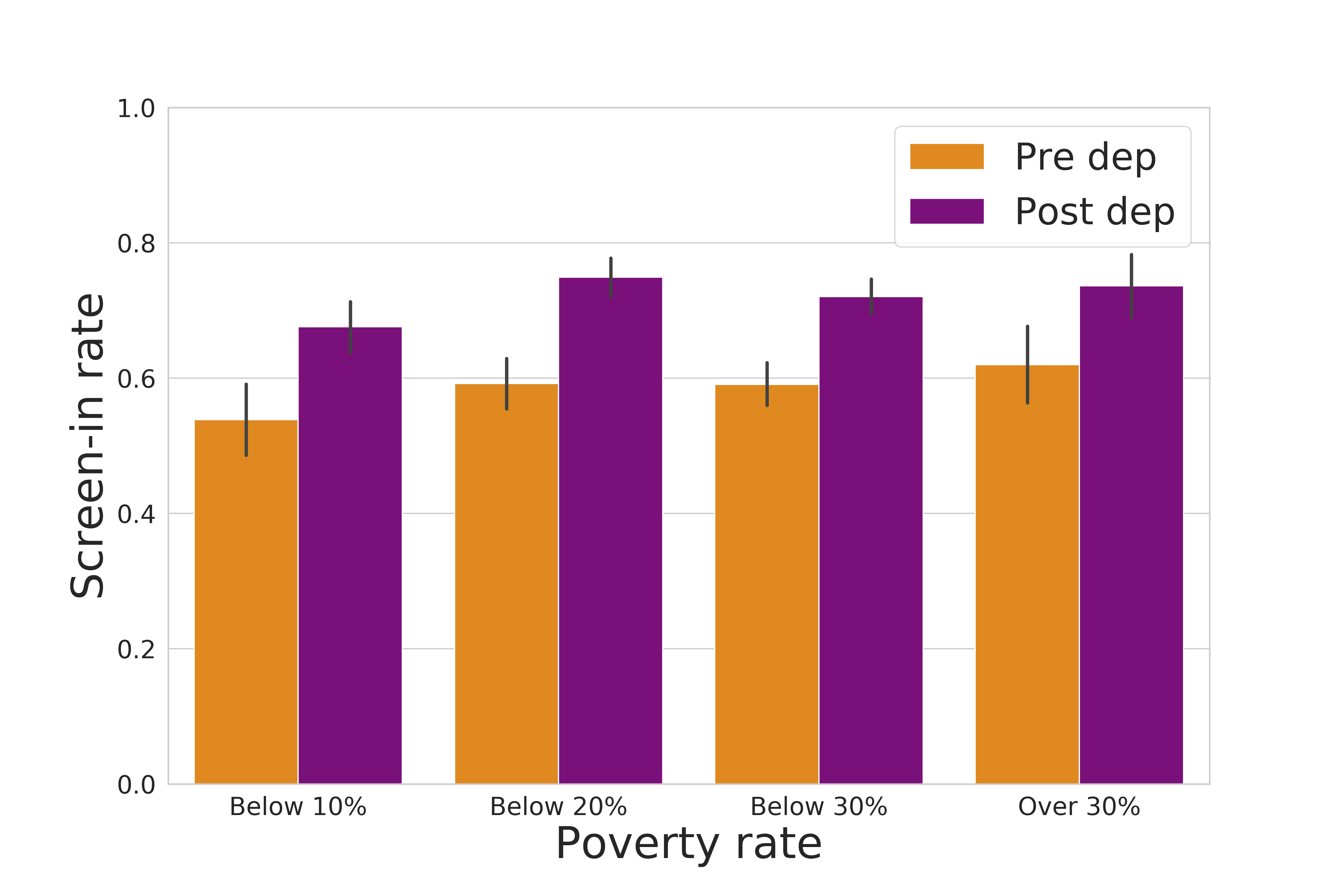}
    \caption{Screen-in rates by fraction of households below the poverty level in the persons' neighbourhood,  pre- and post-deployment, for subset of cases with predicted mandatory screen-in, $M=1$. }
    \label{fig:poverty}
\end{figure}

We also consider disparities across socioeconomic factors. Because the features available for risk assessment rely in part in previous interactions with the public services, one concern that arises is whether differential reliance on the risk assessment tool will disadvantage those who have sought out the government's support in the past. Moreover, and most relevant to the focus of this work, it would be concerning if willingness to override the machine's prediction varied depending on how wealthy the children's family is. If this was the case, we should observe that screen-in rates for cases labeled as $M=1$ decrease as the wealth increased. Figure~\ref{fig:poverty} shows the screen-in rates by poverty rates, defined as \% of households below the poverty level in the persons' neighbourhood. Here, we see that, among assessed mandatory cases, screen in rates do not appear to correlate with neighbourhood poverty levels.

%Black-White disparities? Disparities across poverty levels? 

%Is automation bias only present for a subset of cases, or is adherence to recommendation different across protected groups?

% \subsection{Study limitations}

\section{Discussion}
\label{sec:discussion}

Our analysis studied the adoption of a risk assessment tool that assists call workers tasked with deciding which calls concerning potential child neglect or abuse should be screened in for further investigation. We focused our analysis on investigating whether humans successfully override or ignore erroneous algorithmic scores. We have found that call workers did change their behavior when the tool was deployed, showing partial adherence to the machine's recommendation.  We also found that call workers were less likely to adhere to the recommendations in cases where a technical glitch resulted in an incorrect score being displayed.

When considering how humans make use of recommendations provided by an algorithmic system, algorithm aversion and automation bias can be seen as two ends of a broad spectrum, both of which are undesirable. On one end, algorithm aversion leads the human to completely disregard the machine, even when the recommendation may be providing useful information that would improve the quality of decisions. On the other end, automation bias results in humans blindly following the machine's recommendations, failing to make use of other sources of information and their own judgement to disregard the recommendation when evidence suggests that the machine may be mistaken. The call workers in our study were found to not be at either of these extremes.  Call workers changed their decision-making following the deployment of the tool, but they did not adhere to the tool's recommendations in every instance.  

A clear limitation of our study is that it is retrospective in nature.  However, this retrospective data also offers a rare opportunity to study a phenomenon that would not be ethically feasible to investigate as a randomized field trial.  By virtue of being a study of a real deployed system, our study has the advantage of inherent field validity.  While crowd worker studies and randomized field experiments allow for controlled variation of different factors, even the best conceived randomized trials can fail to have field validity in social policy settings~\cite{Nagin2019}. Moreover, given the high-stakes nature of the task, behaviors displayed by call workers when given hypothetical calls in a lab setting may differ from their behavior when faced with a real allegation of child abuse. Our investigation of the technical glitch demonstrates that certain phenomena are in fact observed to occur when trained experts are making high-stakes decisions in a real world setting.

What contributed to these positive results?  The retrospective nature of our study means that we were unable to perturb different elements of the decision making framework in order to assess their impact on decision outcomes.  However, through our analysis and discussions with jurisdiction staff, we are able to identify certain elements of the deployment setup that we believe likely influenced the observed behavior, which we elaborate on below.  An important direction for future research is to further investigate these and other factors in more controlled settings to gain a better understanding of their influence on commission and omission errors in algorithm assisted decision making.  

A key contributing factor is that throughout the process call workers continued to have access to not only the referral calls but also the administrative data system. This provided a different view of the case than what was being pulled into the risk score calculation. In particular, even when inputs related to past child welfare history were being miscalculated in real time, workers would still have access to the correct information in the data system. In addition to the role of having access to the raw features, and having the time to inspect these, it is particularly important to highlight that call workers had been previously trained to make decisions without the aid of a risk assessment score. Therefore, they had experience in parsing and interpreting the raw data. A question that arises is whether this previous experience had an important role, and whether the same behavior can be expected from call workers who start working after the deployment of the tool. The answer to this question could inform the need for training that teaches experts how to make use of other sources of data appropriately, in order to avoid an over-reliance on algorithmic recommendations. 

Secondly, the risk tool provides workers only with a score, and does not `explain' its predictions, nor does it display values of any of the features involved in the score calculation.  If this additional information had been provided, it is possible that the glitch would have been detected by workers. However, it is also possible that this distillation of the data would have been trusted by workers, who would in turn have been dissuaded from examining the original data.  In the latter case, the workers would likely have been more susceptible to omission errors. 

With regard to design recommendations, the results highlight the fact that humans-in-the-loop can help guard against harmful effects resulting from erroneous algorithmic recommendations. Evidence in other domains has shown that humans-in-the-loop do not always improve the quality of decisions, leading some to call for complete automation of tasks. However, providing humans with autonomy to contradict the machine mitigated the effects of miscalculated scores in the child maltreatment call screening context. Given that technical glitches as the one studied in this paper are always a risk, and that any statistical model will make mistakes, design should focus on augmenting the human's ability to identify and correct mistakes. Future research in controlled settings that evaluates the effect of individual elements of the decision-making pipeline described in this paper could identify specific design practices that effectively strengthen the human's role.

%We note that our findings are significantly different from what has been found in studies of the adoption of risk assessment tools in other domains, such as recidivism prediction for bail decisions. The results presented suggest that in this deployment context, algorithm aversion and automation bias have been at least partially overcome. This finding highlights the fact that experts' adoption of these tools may vary widely across domains and deployment setups. It is important to take this into consideration both when designing algorithms as well as regulatory policy. 

More research is also needed to understand how to develop machine learning decision support tools that provide information helpful to the humans in assessing whether the machine's predictions effectively account for all factors relevant to a case.  This is an especially challenging task in forecasting settings where research shows that human predictions tend to underperform those of statistical models.  Whereas humans (or a consensus of humans) may be expected to identify cases where an image classifier incorrectly tags a dog as a cat, humans are not as reliable in assessing which students will succeed, which children are at risk, or which defendants will appear for court.  We hope that the findings presented in this paper will motivate further work on developing systems that make it easier for humans to assess whether the machine may be making a mistake in the forecasting setting. In particular, many of the decision-pipeline elements likely contributed to workers selectively trusting the machine by questioning its recommendations. Further research is needed to better understand the role that trust and other factors play in the overall success of human-in-the-loop algorithmic systems.  % in the deployment of algorithmic systems will focus on analysing the aspects that led to the successful behavior of the workers that we witnessed in our work. 

\section{ACKNOWLEDGMENTS}
We are grateful to the Hillman Foundation for
funding this research, and to our collaborators at the Allegheny
County Department of Human Services. Thanks also to the anonymous reviewers
for their numerous comments and suggestions that helped to improve the manuscript.  

%There is a critical need to develop a general undestanding of how the model and decision making framework around it can be designed to help overcome the effects of algorithm aversion and automation bias.    %the policy and academic discussions on the topic must account for this. 

%Understanding how call workers are making use of the info to correct the mistakes is part of future work. In particular, identifying what elements of the user interface or deployment context prevents automation bias from taking place. Do they rely on the call, or are they also making direct use of the covariates?

%Analyzing variations across decision-makers. 

%While we have analyzed overall disparities and seen that gaps between blacks and whites is actually reduced, and gaps across income levels do not change, there may be disparities harming smaller subpopulations. E.g. disparities against parents with disabilities. 

%Spillover--incorrect scores on some cases may affect the human trust for other cases. 

% BALANCE COLUMNS
\balance{}

% REFERENCES FORMAT
% References must be the same font size as other body text.
\bibliographystyle{SIGCHI-Reference-Format}
\bibliography{mybib}


\begin{thebibliography}{00}

%%% the Web address.

\ifx \showCODEN    \undefined \def \showCODEN     #1{\unskip}     \fi
\ifx \showDOI      \undefined \def \showDOI       #1{{\tt DOI:}\penalty0{#1}\ }
  \fi
\ifx \showISBNx    \undefined \def \showISBNx     #1{\unskip}     \fi
\ifx \showISBNxiii \undefined \def \showISBNxiii  #1{\unskip}     \fi
\ifx \showISSN     \undefined \def \showISSN      #1{\unskip}     \fi
\ifx \showLCCN     \undefined \def \showLCCN      #1{\unskip}     \fi
\ifx \shownote     \undefined \def \shownote      #1{#1}          \fi
\ifx \showarticletitle \undefined \def \showarticletitle #1{#1}   \fi
\ifx \showURL      \undefined \def \showURL       #1{#1}          \fi

\bibitem{aegisdottir2006meta}
{Stefan{\'\i}a {\AE}gisd{\'o}ttir}, {Michael~J White}, {Paul~M Spengler},
  {Alan~S Maugherman}, {Linda~A Anderson}, {Robert~S Cook}, {Cassandra~N
  Nichols}, {Georgios~K Lampropoulos}, {Blain~S Walker}, {Genna Cohen}, {and}
  {others}. 2006.
\newblock \showarticletitle{The meta-analysis of clinical judgment project:
  Fifty-six years of accumulated research on clinical versus statistical
  prediction}.
\newblock {\em The Counseling Psychologist\/} {34}, 3 (2006), 341--382.
\newblock


\bibitem{evidencealbright}
{Alex Albright}. 2019.
\newblock {\em If You Give a Judge a Risk Score: Evidence from Kentucky Bail
  Decisions}.
\newblock {T}echnical {R}eport. Working paper.
\newblock


\bibitem{bansal2019updates}
{Gagan Bansal}, {Besmira Nushi}, {Ece Kamar}, {Daniel~S Weld}, {Walter~S
  Lasecki}, {and} {Eric Horvitz}. 2019.
\newblock \showarticletitle{Updates in human-ai teams: Understanding and
  addressing the performance/compatibility tradeoff}. In {\em Proceedings of
  the AAAI Conference on Artificial Intelligence}, Vol.~33. 2429--2437.
\newblock


\bibitem{binns2018s}
{Reuben Binns}, {Max Van~Kleek}, {Michael Veale}, {Ulrik Lyngs}, {Jun Zhao},
  {and} {Nigel Shadbolt}. 2018.
\newblock \showarticletitle{'It's Reducing a Human Being to a Percentage':
  Perceptions of Justice in Algorithmic Decisions}. In {\em Proceedings of the
  2018 CHI Conference on Human Factors in Computing Systems}. ACM, 377.
\newblock


\bibitem{brown2019toward}
{Anna Brown}, {Alexandra Chouldechova}, {Emily Putnam-Hornstein}, {Andrew
  Tobin}, {and} {Rhema Vaithianathan}. 2019.
\newblock \showarticletitle{Toward Algorithmic Accountability in Public
  Services: A Qualitative Study of Affected Community Perspectives on
  Algorithmic Decision-making in Child Welfare Services}. In {\em Proceedings
  of the 2019 CHI Conference on Human Factors in Computing Systems}. ACM, 41.
\newblock


\bibitem{bushway2012sentencing}
{Shawn~D Bushway}, {Emily~G Owens}, {and} {Anne~Morrison Piehl}. 2012.
\newblock \showarticletitle{Sentencing guidelines and judicial discretion:
  Quasi-experimental evidence from human calculation errors}.
\newblock {\em Journal of Empirical Legal Studies\/} {9}, 2 (2012), 291--319.
\newblock


\bibitem{caruana2015intelligible}
{Rich Caruana}, {Yin Lou}, {Johannes Gehrke}, {Paul Koch}, {Marc Sturm}, {and}
  {Noemie Elhadad}. 2015.
\newblock \showarticletitle{Intelligible models for healthcare: Predicting
  pneumonia risk and hospital 30-day readmission}. In {\em Proceedings of the
  21th ACM SIGKDD International Conference on Knowledge Discovery and Data
  Mining}. ACM, 1721--1730.
\newblock


\bibitem{chouldechova2018case}
{Alexandra Chouldechova}, {Diana Benavides-Prado}, {Oleksandr Fialko}, {and}
  {Rhema Vaithianathan}. 2018.
\newblock \showarticletitle{A case study of algorithm-assisted decision making
  in child maltreatment hotline screening decisions}. In {\em Conference on
  Fairness, Accountability and Transparency}. 134--148.
\newblock


\bibitem{cohen2019judicial}
{Alma Cohen} {and} {Crystal~S Yang}. 2019.
\newblock \showarticletitle{Judicial politics and sentencing decisions}.
\newblock {\em American Economic Journal: Economic Policy\/} {11}, 1 (2019),
  160--91.
\newblock


\bibitem{cummings2004automation}
{Mary Cummings}. 2004.
\newblock \showarticletitle{Automation bias in intelligent time critical
  decision support systems}. In {\em AIAA 1st Intelligent Systems Technical
  Conference}. 6313.
\newblock


\bibitem{dawes1989clinical}
{Robyn~M Dawes}, {David Faust}, {and} {Paul~E Meehl}. 1989.
\newblock \showarticletitle{Clinical versus actuarial judgment}.
\newblock {\em Science\/} {243}, 4899 (1989), 1668--1674.
\newblock


\bibitem{demichele2018criminal}
{Matthew DeMichele}, {Peter Baumgartner}, {Kelle Barrick}, {Megan Comfort},
  {Samuel Scaggs}, {and} {Shilpi Misra}. 2018.
\newblock \showarticletitle{What do criminal justice professionals think about
  risk assessment at pretrial?}
\newblock {\em Available at SSRN 3168490\/} (2018).
\newblock


\bibitem{dietvorst2015algorithm}
{Berkeley~J Dietvorst}, {Joseph~P Simmons}, {and} {Cade Massey}. 2015.
\newblock \showarticletitle{Algorithm aversion: People erroneously avoid
  algorithms after seeing them err.}
\newblock {\em Journal of Experimental Psychology: General\/} {144}, 1 (2015),
  114.
\newblock


\bibitem{dietvorst2016overcoming}
{Berkeley~J Dietvorst}, {Joseph~P Simmons}, {and} {Cade Massey}. 2016.
\newblock \showarticletitle{Overcoming algorithm aversion: People will use
  imperfect algorithms if they can (even slightly) modify them}.
\newblock {\em Management Science\/} {64}, 3 (2016), 1155--1170.
\newblock


\bibitem{roadblock}
{Jennif Doleac} {and} {Megan Stevenson}. 2018.
\newblock {\em The roadblock to reform}.
\newblock {T}echnical {R}eport. American Constitution Society Research Report.
\newblock


\bibitem{eubanks2018automating}
{Virginia Eubanks}. 2018.
\newblock {\em Automating inequality: How high-tech tools profile, police, and
  punish the poor}.
\newblock St. Martin's Press.
\newblock


\bibitem{goddard2011automation}
{Kate Goddard}, {Abdul Roudsari}, {and} {Jeremy~C Wyatt}. 2011.
\newblock \showarticletitle{Automation bias: a systematic review of frequency,
  effect mediators, and mitigators}.
\newblock {\em Journal of the American Medical Informatics Association\/} {19},
  1 (2011), 121--127.
\newblock


\bibitem{goodwin1999judgmental}
{Paul Goodwin} {and} {Robert Fildes}. 1999.
\newblock \showarticletitle{Judgmental forecasts of time series affected by
  special events: Does providing a statistical forecast improve accuracy?}
\newblock {\em Journal of Behavioral Decision Making\/} {12}, 1 (1999), 37--53.
\newblock


\bibitem{green2019disparate}
{Ben Green} {and} {Yiling Chen}. 2019.
\newblock \showarticletitle{Disparate interactions: An algorithm-in-the-loop
  analysis of fairness in risk assessments}. In {\em Proceedings of the
  Conference on Fairness, Accountability, and Transparency}. ACM, 90--99.
\newblock


\bibitem{grove2000clinical}
{William~M Grove}, {David~H Zald}, {Boyd~S Lebow}, {Beth~E Snitz}, {and} {Chad
  Nelson}. 2000.
\newblock \showarticletitle{Clinical versus mechanical prediction: a
  meta-analysis}.
\newblock {\em Psychological assessment\/} {12}, 1 (2000), 19.
\newblock


\bibitem{hilgard2019learning}
{Sophie Hilgard}, {Nir Rosenfeld}, {Mahzarin~R Banaji}, {Jack Cao}, {and}
  {David~C Parkes}. 2019.
\newblock \showarticletitle{Learning Representations by Humans, for Humans}.
\newblock {\em arXiv preprint arXiv:1905.12686\/} (2019).
\newblock


\bibitem{jhaver2019does}
{Shagun Jhaver}, {Amy Bruckman}, {and} {Eric Gilbert}. 2019.
\newblock \showarticletitle{Does transparency in moderation really matter? User
  behavior after content removal explanations on reddit}.
\newblock {\em Proceedings of the ACM on Human-Computer Interaction\/} {3},
  CSCW (2019), 1--27.
\newblock


\bibitem{kehl2017algorithms}
{Danielle~Leah Kehl} {and} {Samuel~Ari Kessler}. 2017.
\newblock \showarticletitle{Algorithms in the criminal justice system:
  Assessing the use of risk assessments in sentencing}.
\newblock  (2017).
\newblock


\bibitem{kizilcec2016much}
{Ren{\'e}~F Kizilcec}. 2016.
\newblock \showarticletitle{How much information?: Effects of transparency on
  trust in an algorithmic interface}. In {\em Proceedings of the 2016 CHI
  Conference on Human Factors in Computing Systems}. ACM, 2390--2395.
\newblock


\bibitem{kleinberg2017human}
{Jon Kleinberg}, {Himabindu Lakkaraju}, {Jure Leskovec}, {Jens Ludwig}, {and}
  {Sendhil Mullainathan}. 2017.
\newblock \showarticletitle{Human decisions and machine predictions}.
\newblock {\em The quarterly journal of economics\/} {133}, 1 (2017), 237--293.
\newblock


\bibitem{kube2019allocating}
{Amanda Kube}, {Sanmay Das}, {and} {Patrick~J Fowler}. 2019.
\newblock \showarticletitle{Allocating interventions based on predicted
  outcomes: A case study on homelessness services}. In {\em Proceedings of the
  AAAI Conference on Artificial Intelligence}.
\newblock


\bibitem{lakkaraju2019fool}
{Himabindu Lakkaraju} {and} {Osbert Bastani}. 2019.
\newblock \showarticletitle{" How do I fool you?": Manipulating User Trust via
  Misleading Black Box Explanations}.
\newblock {\em arXiv preprint arXiv:1911.06473\/} (2019).
\newblock


\bibitem{lee2004trust}
{John~D Lee} {and} {Katrina~A See}. 2004.
\newblock \showarticletitle{Trust in automation: Designing for appropriate
  reliance}.
\newblock {\em Human factors\/} {46}, 1 (2004), 50--80.
\newblock


\bibitem{lim1995judgemental}
{Joa~Sang Lim} {and} {Marcus O'Connor}. 1995.
\newblock \showarticletitle{Judgemental adjustment of initial forecasts: Its
  effectiveness and biases}.
\newblock {\em Journal of Behavioral Decision Making\/} {8}, 3 (1995),
  149--168.
\newblock


\bibitem{madras2018predict}
{David Madras}, {Toni Pitassi}, {and} {Richard Zemel}. 2018.
\newblock \showarticletitle{Predict responsibly: improving fairness and
  accuracy by learning to defer}. In {\em Advances in Neural Information
  Processing Systems}. 6147--6157.
\newblock


\bibitem{marten2004computer}
{Katharina Marten}, {Tobias Seyfarth}, {Florian Auer}, {Edzard Wiener},
  {Andreas Grillh{\"o}sl}, {Silvia Obenauer}, {Ernst~J Rummeny}, {and}
  {Christoph Engelke}. 2004.
\newblock \showarticletitle{Computer-assisted detection of pulmonary nodules:
  performance evaluation of an expert knowledge-based detection system in
  consensus reading with experienced and inexperienced chest radiologists}.
\newblock {\em European radiology\/} {14}, 10 (2004), 1930--1938.
\newblock


\bibitem{meehl1954clinical}
{Paul~E Meehl}. 1954.
\newblock \showarticletitle{Clinical versus statistical prediction: A
  theoretical analysis and a review of the evidence}. In {\em In Proceedings of
  the 1955 Invitational Conference on Testing Problems}. University of
  Minnesota Press, 136--141.
\newblock


\bibitem{moray2000adaptive}
{Neville Moray}, {Toshiyuki Inagaki}, {and} {Makoto Itoh}. 2000.
\newblock \showarticletitle{Adaptive automation, trust, and self-confidence in
  fault management of time-critical tasks.}
\newblock {\em Journal of experimental psychology: Applied\/} {6}, 1 (2000),
  44.
\newblock


\bibitem{mosier1998automation}
{Kathleen~L Mosier}, {Melisa Dunbar}, {Lori McDonnell}, {Linda~J Skitka}, {Mark
  Burdick}, {and} {Bonnie Rosenblatt}. 1998a.
\newblock \showarticletitle{Automation bias and errors: Are teams better than
  individuals?}. In {\em Proceedings of the Human Factors and Ergonomics
  Society Annual Meeting}, Vol.~42. SAGE Publications Sage CA: Los Angeles, CA,
  201--205.
\newblock


\bibitem{mosier1998}
{Kathleen~L. Mosier}, {Linda~J. Skitka}, {Susan Heers}, {and} {Mark Burdick}.
  1998b.
\newblock \showarticletitle{Automation Bias: Decision Making and Performance in
  High-Tech Cockpits}.
\newblock {\em The International Journal of Aviation Psychology\/} {8}, 1
  (1998), 47--63.
\newblock
\showDOI{%
\url{http://dx.doi.org/10.1207/s15327108ijap0801\_3}}


\bibitem{Nagin2019}
{Daniel~S. Nagin} {and} {Robert~J. Sampson}. 2019.
\newblock \showarticletitle{The Real Gold Standard: Measuring Counterfactual
  Worlds That Matter Most to Social Science and Policy}.
\newblock {\em Annual Review of Criminology\/} {2}, 1 (2019), 123--145.
\newblock
\showDOI{%
\url{http://dx.doi.org/10.1146/annurev-criminol-011518-024838}}


\bibitem{nourani2019effects}
{Mahsan Nourani}, {Samia Kabir}, {Sina Mohseni}, {and} {Eric~D Ragan}. 2019.
\newblock \showarticletitle{The Effects of Meaningful and Meaningless
  Explanations on Trust and Perceived System Accuracy in Intelligent Systems}.
  In {\em Proceedings of the AAAI Conference on Human Computation and
  Crowdsourcing}, Vol.~7. 97--105.
\newblock


\bibitem{afst}
{Allegheny County~Department of Health} {and} {Human Services}. n.d.
\newblock Allegheny Family Screening Tool.
\newblock   (n.d.).
\newblock
\showURL{%
\url{https://www.alleghenycounty.us/Human-Services/News-Events/Accomplishments/Allegheny-Family-Screening-Tool.aspx}}


\bibitem{rader2018explanations}
{Emilee Rader}, {Kelley Cotter}, {and} {Janghee Cho}. 2018.
\newblock \showarticletitle{Explanations as mechanisms for supporting
  algorithmic transparency}. In {\em Proceedings of the 2018 CHI Conference on
  Human Factors in Computing Systems}. ACM, 103.
\newblock


\bibitem{raghu2019algorithmic}
{Maithra Raghu}, {Katy Blumer}, {Greg Corrado}, {Jon Kleinberg}, {Ziad
  Obermeyer}, {and} {Sendhil Mullainathan}. 2019.
\newblock \showarticletitle{The algorithmic automation problem: Prediction,
  triage, and human effort}.
\newblock {\em arXiv preprint arXiv:1903.12220\/} (2019).
\newblock


\bibitem{sarter2001supporting}
{Nadine~B Sarter} {and} {Beth Schroeder}. 2001.
\newblock \showarticletitle{Supporting decision making and action selection
  under time pressure and uncertainty: The case of in-flight icing}.
\newblock {\em Human factors\/} {43}, 4 (2001), 573--583.
\newblock


\bibitem{skeem2019impact}
{Jennifer~L Skeem}, {Nicholas Scurich}, {and} {John Monahan}. 2019.
\newblock \showarticletitle{Impact of Risk Assessment on Judges’ Fairness in
  Sentencing Relatively Poor Defendants}.
\newblock {\em Virginia Public Law and Legal Theory Research Paper\/} 2019-02
  (2019).
\newblock


\bibitem{skitka2000automation}
{Linda~J Skitka}, {Kathleen~L Mosier}, {Mark Burdick}, {and} {Bonnie
  Rosenblatt}. 2000.
\newblock \showarticletitle{Automation bias and errors: are crews better than
  individuals?}
\newblock {\em The International journal of aviation psychology\/} {10}, 1
  (2000), 85--97.
\newblock


\bibitem{sloan2018effect}
{CarlyWill Sloan}, {George Naufal}, {and} {Heather Caspers}. 2018.
\newblock \showarticletitle{The Effect of Risk Assessment Scores on Judicial
  Behavior and Defendant Outcomes}.
\newblock  (2018).
\newblock


\bibitem{smith2012predictive}
{Vernon~C Smith}, {Adam Lange}, {and} {Daniel~R Huston}. 2012.
\newblock \showarticletitle{Predictive modeling to forecast student outcomes
  and drive effective interventions in online community college courses.}
\newblock {\em Journal of Asynchronous Learning Networks\/} {16}, 3 (2012),
  51--61.
\newblock


\bibitem{stevenson2018assessing}
{Megan Stevenson}. 2018.
\newblock \showarticletitle{Assessing risk assessment in action}.
\newblock {\em Minn. L. Rev.\/}  {103} (2018), 303.
\newblock


\bibitem{tan2018investigating}
{Sarah Tan}, {Julius Adebayo}, {Kori Inkpen}, {and} {Ece Kamar}. 2018.
\newblock \showarticletitle{Investigating Human+ Machine Complementarity for
  Recidivism Predictions}.
\newblock {\em arXiv preprint arXiv:1808.09123\/} (2018).
\newblock


\bibitem{van2019crowdsourcing}
{Niels van Berkel}, {Jorge Goncalves}, {Danula Hettiachchi}, {Senuri
  Wijenayake}, {Ryan~M Kelly}, {and} {Vassilis Kostakos}. 2019.
\newblock \showarticletitle{Crowdsourcing Perceptions of Fair Predictors for
  Machine Learning: A Recidivism Case Study}.
\newblock {\em Proceedings of the ACM on Human-Computer Interaction\/} {3},
  CSCW (2019), 1--21.
\newblock


\bibitem{yeomans2017making}
{Michael Yeomans}, {Anuj Shah}, {Sendhil Mullainathan}, {and} {Jon Kleinberg}.
  2017.
\newblock \showarticletitle{Making sense of recommendations}.
\newblock {\em Journal of Behavioral Decision Making\/} (2017).
\newblock


\bibitem{yin2019understanding}
{Ming Yin}, {Jennifer Wortman~Vaughan}, {and} {Hanna Wallach}. 2019.
\newblock \showarticletitle{Understanding the Effect of Accuracy on Trust in
  Machine Learning Models}. In {\em Proceedings of the 2019 CHI Conference on
  Human Factors in Computing Systems}. ACM, 279.
\newblock


\bibitem{yu2016trust}
{Kun Yu}, {Shlomo Berkovsky}, {Dan Conway}, {Ronnie Taib}, {Jianlong Zhou},
  {and} {Fang Chen}. 2016.
\newblock \showarticletitle{Trust and reliance based on system accuracy}. In
  {\em Proceedings of the 2016 Conference on User Modeling Adaptation and
  Personalization}. ACM, 223--227.
\newblock


\bibitem{yu2017user}
{Kun Yu}, {Shlomo Berkovsky}, {Ronnie Taib}, {Dan Conway}, {Jianlong Zhou},
  {and} {Fang Chen}. 2017.
\newblock \showarticletitle{User trust dynamics: An investigation driven by
  differences in system performance}. In {\em Proceedings of the 22nd
  International Conference on Intelligent User Interfaces}. ACM, 307--317.
\newblock


\end{thebibliography}

\end{document}